\pdfoutput=1
\documentclass[12pt,a4paper]{article}
\usepackage{ifthen} % for conditional statements
\newboolean{pdflatex}
\setboolean{pdflatex}{true} % False for eps figures 

\newboolean{articletitles}
\setboolean{articletitles}{true} % False removes titles in references

\newboolean{uprightparticles}
\setboolean{uprightparticles}{false} %True for upright particle symbols

\usepackage{rotating}
\usepackage{amsmath}
\usepackage{float}
\usepackage{booktabs}

\def\paperauthors{LHCb collaboration} 
\def\paperasciititle{Measurement of Z boson production cross-section in pp collisions at 5.02 TeV} 
\def\paperkeywords{{High Energy Physics}, {LHCb}}
\def\papercopyright{\the\year\ CERN for the benefit of the LHCb collaboration} 
\def\paperlicence{CC BY 4.0 licence}
\def\paperlicenceurl{https://creativecommons.org/licenses/by/4.0/}

\def\Zmm     {\decay{\Z}{\mumu}} 
\def\Ztt     {\decay{\Z}{\tautau}}
\def\zpt     {\ensuremath{p_{\mathrm{T}}^{Z}}\xspace}
\def\zy      {\ensuremath{y^{{Z}}}\xspace}
\def\ww      {W^{+}W^{-}}
\def\wz      {W^{\pm}\Z}
\def\zz      {\Z\Z}
\def\phistar {\ensuremath{\phi_{\eta}^{*}}\xspace}
\def\totxsec {\ensuremath{39.6 \pm 0.7 \stat \pm 0.6 \syst \pm 0.8 \lumi \, \pb}}
\def\NMFf    {\ensuremath{1.2^{+0.5}_{-0.3}\stat \pm 0.1\syst}\xspace}
\def\NMFb    {\ensuremath{3.6^{+1.6}_{-0.9}\stat \pm 0.2\syst}\xspace}
\def\XSecFor {\ensuremath{13.5^{+5.4}_{-4.0}\stat \pm 1.2\syst\nb}\xspace}
\def\XSecBack {\ensuremath{10.7^{+8.4}_{-5.1}\stat \pm 1.0\syst\nb}\xspace}

\def\pPb   {\ensuremath{\proton\mathrm{Pb}}\xspace}

\def\rpa   {\ensuremath{R_{\proton \mathrm{Pb}}}\xspace}
\def\kpa   {\ensuremath{k_{\proton \mathrm{Pb}}}\xspace}

\def\murapstar {\ensuremath{y^{*}_{\mu}}\xspace}

\usepackage[top=1in, bottom=1.25in, left=1in, right=1in]{geometry}

\usepackage{microtype}
\usepackage{lineno}  % for line numbering during review
\usepackage{xspace} % To avoid problems with missing or double spaces after
\usepackage{caption} %these three command get the figure and table captions automatically small

\usepackage{graphicx}  % to include figures (can also use other packages)
\usepackage{color}
\usepackage{colortbl}
\graphicspath{{./figs/}} % Make Latex search fig subdir for figures
\usepackage{amsmath} % Adds a large collection of math symbols
\usepackage{amssymb}
\usepackage{amsfonts}
\usepackage{upgreek} % Adds in support for greek letters in roman typeset

\newcommand*\patchAmsMathEnvironmentForLineno[1]{%
\expandafter\let\csname old#1\expandafter\endcsname\csname #1\endcsname
\expandafter\let\csname oldend#1\expandafter\endcsname\csname
end#1\endcsname
 \renewenvironment{#1}%
   {\linenomath\csname old#1\endcsname}%
   {\csname oldend#1\endcsname\endlinenomath}%
}
\newcommand*\patchBothAmsMathEnvironmentsForLineno[1]{%
  \patchAmsMathEnvironmentForLineno{#1}%
  \patchAmsMathEnvironmentForLineno{#1*}%
}
\AtBeginDocument{%
\patchBothAmsMathEnvironmentsForLineno{equation}%
\patchBothAmsMathEnvironmentsForLineno{align}%
\patchBothAmsMathEnvironmentsForLineno{flalign}%
\patchBothAmsMathEnvironmentsForLineno{alignat}%
\patchBothAmsMathEnvironmentsForLineno{gather}%
\patchBothAmsMathEnvironmentsForLineno{multline}%
\patchBothAmsMathEnvironmentsForLineno{eqnarray}%
}

\usepackage{hyperxmp}

\usepackage[pdftex,
            pdfauthor={\paperauthors},
            pdftitle={\paperasciititle},
            pdfkeywords={\paperkeywords},
            pdfcopyright={Copyright (C) \papercopyright},
            pdflicenseurl={\paperlicenceurl}]{hyperref}
\usepackage[colorinlistoftodos,textsize=scriptsize]{todonotes}

\usepackage[bottom,flushmargin,hang,multiple]{footmisc}

\usepackage[all]{hypcap} % Internal hyperlinks to floats.

\usepackage{xspace} 
\usepackage{upgreek}

\def\lhcb   {\mbox{LHCb}\xspace}

\def\MagUp {\mbox{\em Mag\kern -0.05em Up}\xspace}

\ifthenelse{\boolean{uprightparticles}}%
{

 \def\Pmu         {\ensuremath{\upmu}\xspace}

 \def\Ptau        {\ensuremath{\uptau}\xspace}

 \def\PDelta      {\ensuremath{\Delta}\xspace}                 
 \def\PXi         {\ensuremath{\Xi}\xspace}                 
 \def\PLambda     {\ensuremath{\Lambda}\xspace}                 
 \def\PSigma      {\ensuremath{\Sigma}\xspace}                 
 \def\POmega      {\ensuremath{\Omega}\xspace}                 
 \def\PUpsilon    {\ensuremath{\Upsilon}\xspace}
 \let\oldPi\Pi
 \def\PPi         {\ensuremath{\oldPi}\xspace}

 \def\PB      {\ensuremath{\mathrm{B}}\xspace}                 
                  
 \def\PD      {\ensuremath{\mathrm{D}}\xspace}

 \def\PK      {\ensuremath{\mathrm{K}}\xspace}

 \def\PW      {\ensuremath{\mathrm{W}}\xspace}

 \def\PZ      {\ensuremath{\mathrm{Z}}\xspace}                 
                  
 \def\Pb      {\ensuremath{\mathrm{b}}\xspace}                 
 \def\Pc      {\ensuremath{\mathrm{c}}\xspace}

 \def\Pi      {\ensuremath{\mathrm{i}}\xspace}

 \def\Pp      {\ensuremath{\mathrm{p}}\xspace}

 \def\Ps      {\ensuremath{\mathrm{s}}\xspace}                 
 \def\Pt      {\ensuremath{\mathrm{t}}\xspace}

 \def\thebaroffset{0.0em}
}
{

 \def\Pmu         {\ensuremath{\mu}\xspace}

 \def\Ptau        {\ensuremath{\tau}\xspace}

 \mathchardef\PDelta="7101
 \mathchardef\PXi="7104
 \mathchardef\PLambda="7103
 \mathchardef\PSigma="7106
 \mathchardef\POmega="710A
 \mathchardef\PUpsilon="7107
 \mathchardef\PPi="7105
                  
 \def\PB      {\ensuremath{B}\xspace}                 
                  
 \def\PD      {\ensuremath{D}\xspace}

 \def\PK      {\ensuremath{K}\xspace}

 \def\PW      {\ensuremath{W}\xspace}

 \def\PZ      {\ensuremath{Z}\xspace}                 
                  
 \def\Pb      {\ensuremath{b}\xspace}                 
 \def\Pc      {\ensuremath{c}\xspace}

 \def\Pi      {\ensuremath{i}\xspace}

 \def\Pp      {\ensuremath{p}\xspace}

 \def\Ps      {\ensuremath{s}\xspace}                 
 \def\Pt      {\ensuremath{t}\xspace}

 \def\thebaroffset{0.18em}
}
\newcommand{\offsetoverline}[2][\thebaroffset]{\kern #1\overline{\kern -#1 #2}}%

\makeatletter
\ifcase \@ptsize \relax% 10pt
  \newcommand{\miniscule}{\@setfontsize\miniscule{4}{5}}% \tiny: 5/6
\or% 11pt
  \newcommand{\miniscule}{\@setfontsize\miniscule{5}{6}}% \tiny: 6/7
\or% 12pt
  \newcommand{\miniscule}{\@setfontsize\miniscule{5}{6}}% \tiny: 6/7
\fi
\makeatother

\DeclareRobustCommand{\optbar}[1]{\shortstack{{\miniscule (\rule[.5ex]{1.25em}{.18mm})}
  \\ [-.7ex] $#1$}}

\def\mup        {{\ensuremath{\Pmu^+}}\xspace}
\def\mun        {{\ensuremath{\Pmu^-}}\xspace} % muon negative (\mum is taken)

\def\mumu       {{\ensuremath{\Pmu^+\Pmu^-}}\xspace}

\def\tautau     {{\ensuremath{\Ptau^+\Ptau^-}}\xspace}

\def\W      {{\ensuremath{\PW}}\xspace}

\def\Z      {{\ensuremath{\PZ}}\xspace}

\def\squark    {{\ensuremath{\Ps}}\xspace}

\def\cquark    {{\ensuremath{\Pc}}\xspace}
\def\cquarkbar {{\ensuremath{\overline \cquark}}\xspace}
\def\ccbar     {{\ensuremath{\cquark\cquarkbar}}\xspace}
\def\bquark    {{\ensuremath{\Pb}}\xspace}
\def\bquarkbar {{\ensuremath{\overline \bquark}}\xspace}
\def\bbbar     {{\ensuremath{\bquark\bquarkbar}}\xspace}
\def\tquark    {{\ensuremath{\Pt}}\xspace}
\def\tquarkbar {{\ensuremath{\overline \tquark}}\xspace}
\def\ttbar     {{\ensuremath{\tquark\tquarkbar}}\xspace}

\def\KorKbar {\kern \thebaroffset\optbar{\kern -\thebaroffset \PK}{}\xspace}

\def\D       {{\ensuremath{\PD}}\xspace}

\def\DorDbar {\kern \thebaroffset\optbar{\kern -\thebaroffset \PD}\xspace}

\def\Dp      {{\ensuremath{\D^+}}\xspace}
\def\Dm      {{\ensuremath{\D^-}}\xspace}

\def\DpDm    {\ensuremath{\Dp {\kern -0.16em \Dm}}\xspace}

\def\B       {{\ensuremath{\PB}}\xspace}

\def\BorBbar {\kern \thebaroffset\optbar{\kern -\thebaroffset \PB}\xspace}

\def\Bd      {{\ensuremath{\B^0}}\xspace}

\def\BdorBdbar {\kern \thebaroffset\optbar{\kern -\thebaroffset \Bd}\xspace}

\def\Bs      {{\ensuremath{\B^0_\squark}}\xspace}

\def\BsorBsbar {\kern \thebaroffset\optbar{\kern -\thebaroffset \Bs}\xspace}

\def\Y#1S{\ensuremath{\PUpsilon{(#1S)}}\xspace}

\def\proton      {{\ensuremath{\Pp}}\xspace}

\def\LorLbar     {\kern \thebaroffset\optbar{\kern -\thebaroffset \PLambda}\xspace}

\newcommand{\decay}[2]{\ensuremath{#1\!\to #2}\xspace} 

\def\to                 {\ensuremath{\rightarrow}\xspace}

\def\AT#1     {\ensuremath{A_{\mathrm{T}}^{#1}}\xspace}           % 2

\def\C#1      {\ensuremath{\mathcal{C}_{#1}}\xspace}                       % 9
\def\Cp#1     {\ensuremath{\mathcal{C}_{#1}^{'}}\xspace}                    % 7
\def\Ceff#1   {\ensuremath{\mathcal{C}_{#1}^{\mathrm{(eff)}}}\xspace}        % 9  
\def\Cpeff#1  {\ensuremath{\mathcal{C}_{#1}^{'\mathrm{(eff)}}}\xspace}       % 7
\def\Ope#1    {\ensuremath{\mathcal{O}_{#1}}\xspace}                       % 2
\def\Opep#1   {\ensuremath{\mathcal{O}_{#1}^{'}}\xspace}                    % 7

\newcommand{\nospaceunit}[1]{\ensuremath{\text{#1}}}       
\newcommand{\aunit}[1]{\ensuremath{\text{\,#1}}}       
\newcommand{\tev}{\aunit{Te\kern -0.1em V}\xspace}
\newcommand{\gev}{\aunit{Ge\kern -0.1em V}\xspace}
\newcommand{\mev}{\aunit{Me\kern -0.1em V}\xspace}
\newcommand{\kev}{\aunit{ke\kern -0.1em V}\xspace}
\newcommand{\ev}{\aunit{e\kern -0.1em V}\xspace}
 
\newcommand{\mevc}{\ensuremath{\aunit{Me\kern -0.1em V\!/}c}\xspace}
\newcommand{\gevc}{\ensuremath{\aunit{Ge\kern -0.1em V\!/}c}\xspace}
\newcommand{\mevcc}{\ensuremath{\aunit{Me\kern -0.1em V\!/}c^2}\xspace}
\newcommand{\gevcc}{\ensuremath{\aunit{Ge\kern -0.1em V\!/}c^2}\xspace}
\def\mum  {\ensuremath{\,\upmu\nospaceunit{m}}\xspace}

\def\nb {\aunit{nb}\xspace}

\def\pb {\aunit{pb}\xspace}
\def\invpb {\ensuremath{\pb^{-1}}\xspace}

\newcommand{\stat}{\aunit{(stat)}\xspace}
\newcommand{\syst}{\aunit{(syst)}\xspace}
\newcommand{\lumi}{\aunit{(lumi)}\xspace}

\newcommand{\chisq}{\ensuremath{\chi^2}\xspace}

\def\gsim{{~\raise.15em\hbox{$>$}\kern-.85em
          \lower.35em\hbox{$\sim$}~}\xspace}
\def\lsim{{~\raise.15em\hbox{$<$}\kern-.85em
          \lower.35em\hbox{$\sim$}~}\xspace}

\def\sqs   {\ensuremath{\protect\sqrt{s}}\xspace}

\def\pt         {\ensuremath{p_{\mathrm{T}}}\xspace}

\def\ptot       {\ensuremath{p}\xspace}

\newcommand{\lum} {\ensuremath{\mathcal{L}}\xspace}
\def\geant      {\mbox{\textsc{Geant4}}\xspace}

\def\photos     {\mbox{\textsc{Photos}}\xspace}
\def\powheg     {\mbox{\textsc{Powheg}}\xspace}
\def\pythia     {\mbox{\textsc{Pythia}}\xspace}
\def\resbos     {\mbox{\textsc{ResBos}}\xspace}

\def\tell1  {TELL1\xspace}
\def\ukl1   {UKL1\xspace}

\newcommand{\lhcborcid}[1]{\href{https://orcid.org/#1}{\hspace*{0.1em}\raisebox{-0.45ex}{\includegraphics[width=1em]{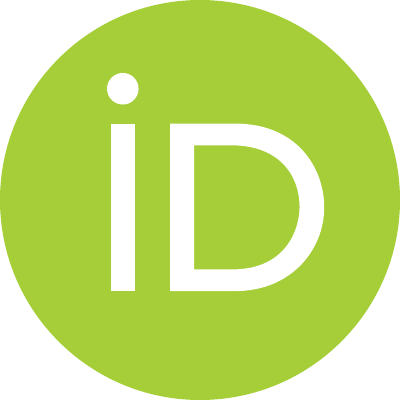}}}}

\usepackage{cite} % Allows for ranges in citations
\usepackage{mciteplus}
\usepackage{longtable}
\begin{document}

%%%%%%%%%%%%%%%%%%%%%%%%%
\renewcommand{\thefootnote}{\fnsymbol{footnote}}
\setcounter{footnote}{1}

% %%%%%%% CHOOSE TITLE PAGE--------
%\onecolumn
%\twocolumn
% ===============================================================================
% Purpose: LHCb-PAPER journal paper title page template
% Author: 
% Created on: 2010-09-25
% ===============================================================================

%%%%%%%%%%%%%%%%%%%%%%%%%
%%%%%  TITLE PAGE  %%%%%%
%%%%%%%%%%%%%%%%%%%%%%%%%
\begin{titlepage}
\pagenumbering{roman}

% Header ---------------------------------------------------
\vspace*{-1.5cm}
\centerline{\large EUROPEAN ORGANIZATION FOR NUCLEAR RESEARCH (CERN)}
\vspace*{1.5cm}
\noindent
\begin{tabular*}{\linewidth}{lc@{\extracolsep{\fill}}r@{\extracolsep{0pt}}}
\ifthenelse{\boolean{pdflatex}}% Logo format choice
{\vspace*{-1.5cm}\mbox{\!\!\!\includegraphics[width=.14\textwidth]{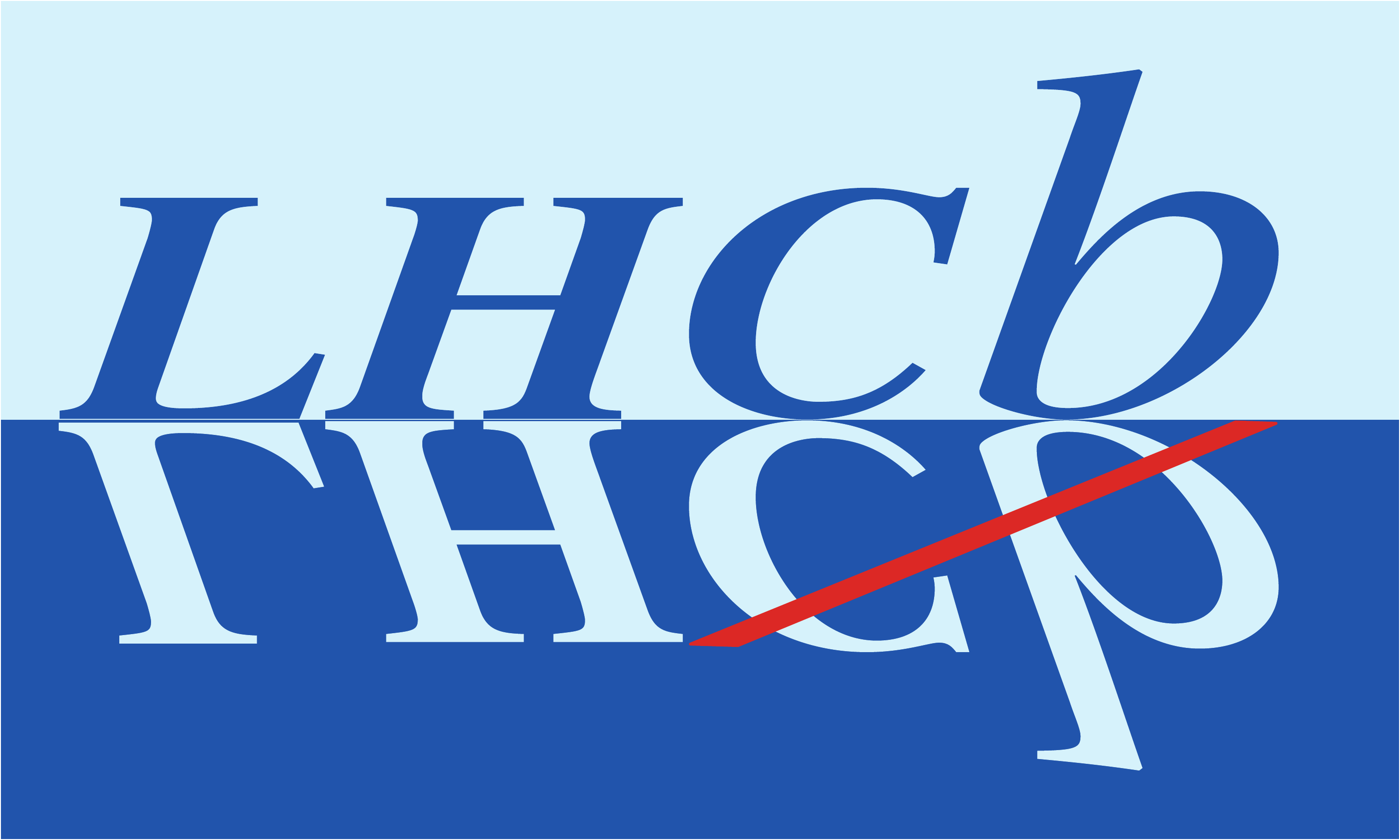}} & &}%
{\vspace*{-1.2cm}\mbox{\!\!\!\includegraphics[width=.12\textwidth]{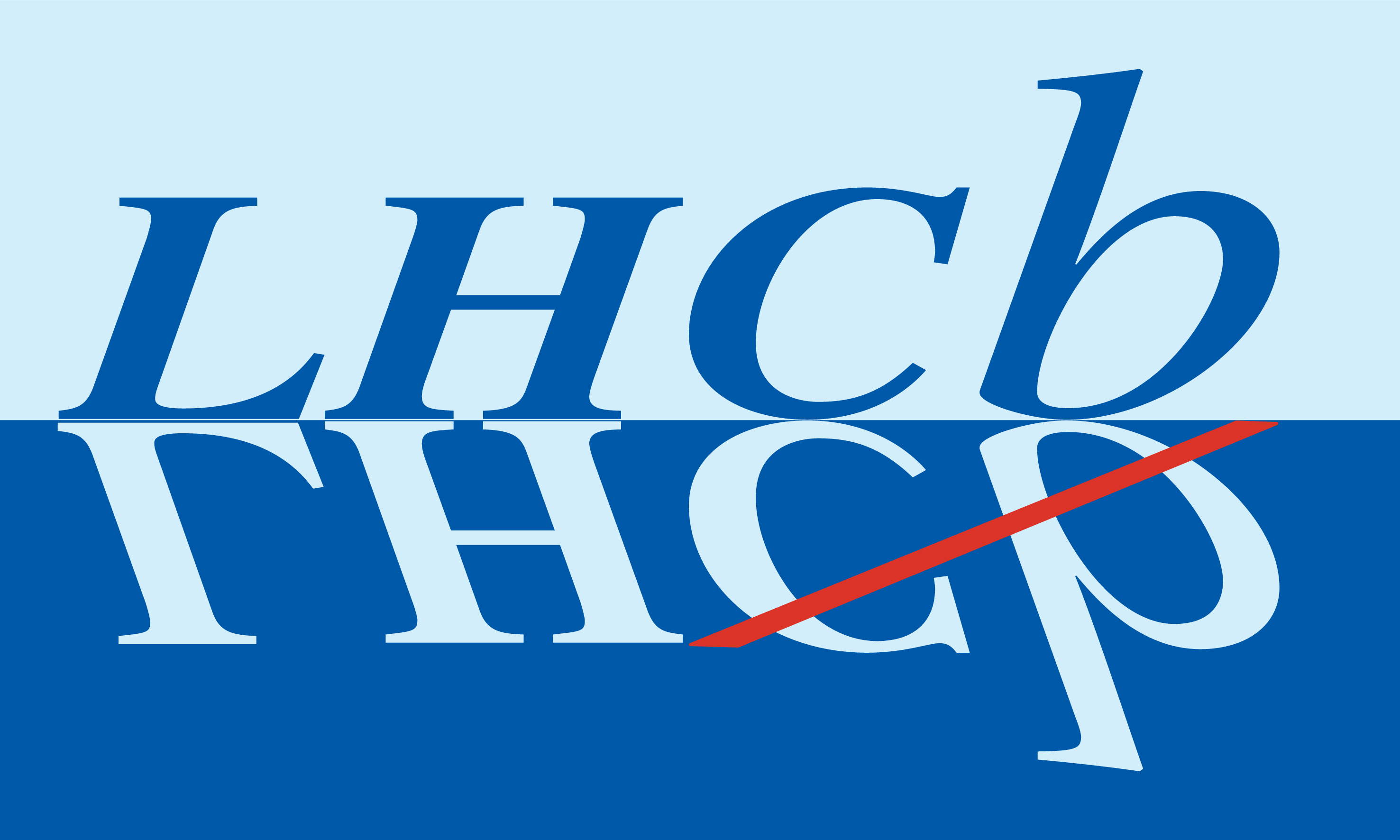}} & &}%
\\
 & & CERN-EP-2023-141 \\  % ID 
 & & LHCb-PAPER-2023-010 \\  % ID 
& & February  21,  2024
% & & \today
\end{tabular*}

\vspace*{4.0cm}

% Title --------------------------------------------------
{\normalfont\bfseries\boldmath\huge
\begin{center}
% DO NOT EDIT HERE. Instead edit macro in main.tex to keep metadata correct
Measurement of the \Z boson production cross-section in $pp$ collisions at \sqs $=5.02\tev$
% Measurement of $Z$ boson production cross-section in $pp$ collisions at $\sqrt{s} = 5.02$ TeV
%\papertitle
\end{center}
}

\vspace*{.5cm}
%\vspace*{2.0cm}

% Authors -------------------------------------------------
\begin{center}
%In the footnote, replace 'paper' by 'Letter' in case of submission to PRL or PLB 
% Edit macro in main.tex to keep metadata correct
\paperauthors\footnote{Authors are listed at the end of this paper.}
\end{center}

\vspace{\fill}

% Abstract -----------------------------------------------
\begin{abstract}
\noindent
The first measurement of the \Z boson production cross-section at centre-of-mass energy $\sqs = 5.02\tev$ in the forward region is reported, using $pp$ collision data collected by the \lhcb experiment in year 2017, 
corresponding to an integrated luminosity of $100 \pm 2 \invpb$. 
The production cross-section is measured for final-state muons in the pseudorapidity range $2.0<\eta<4.5$ with transverse momentum $\pt>20\gevc$. The integrated cross-section is determined to be 
\begin{equation}
\sigma_{\Zmm} = \totxsec
\nonumber
\end{equation}
for the di-muon invariant mass in the range $60<M_{\mu\mu}<120\gevcc$.
This result and the differential cross-section results are in good agreement with theoretical predictions at next-to-next-to-leading order in the strong coupling constant.

Based on a previous \lhcb measurement of the \Z boson production cross-section in \pPb collisions at $\sqrt{s_{NN}}=5.02\tev$, 
the nuclear modification factor \rpa is measured for the first time at this energy. 
The measured values are $\mbox{\NMFf}$ in the forward region ($1.53<\murapstar<4.03$) and $\mbox{\NMFb}$ in the backward region ($-4.97<\murapstar<-2.47$), 
where $y^*_{\mu}$ represents the muon rapidity in the centre-of-mass frame.
\end{abstract}
\vspace*{1.0cm}

\begin{center}
  Published in JHEP 02 (2024) 070
  % Phys.~Rev.~D 
%  Phys.~Rev.~Lett. 
%  Phys.~Lett.~B /
 % Eur.~Phys.~J.~C /
  %  Nucl.~Phys.~B /
%  Chin.~Phys.~C 
%  Nature~Physics /
 % sciPost~Physics /
 % J. Instr. /
 % Instruments 
\end{center}

\vspace{\fill}

{\footnotesize 
% Edit macro in main.tex to keep metadata correct
\centerline{\copyright~\papercopyright. \href{\paperlicenceurl}{\paperlicence}.}}
\vspace*{2mm}

\end{titlepage}

%%%%%%%%%%%%%%%%%%%%%%%%%%%%%%%%
%%%%%  EOD OF TITLE PAGE  %%%%%%
%%%%%%%%%%%%%%%%%%%%%%%%%%%%%%%%

%  empty page follows the title page ----
\newpage
\setcounter{page}{2}
\mbox{~}
%\newpage

% %%%%%%% ---------

\renewcommand{\thefootnote}{\arabic{footnote}}
\setcounter{footnote}{0}

%%%% Uncomment if desired
%\tableofcontents
% \linenumbers

\cleardoublepage
\pagestyle{plain} % restore page numbers for the main text
\setcounter{page}{1}
\pagenumbering{arabic}

%%% introduction
\section{Introduction}
\label{sec:introduction}
%Measurements of 
The $pp \rightarrow\Z \rightarrow \mu^{+}\mu^{-}$ process\footnote{The production process should be interpreted as $pp \to Z/\gamma^{*} \to \mup\mun$ in the strict sense. In this article, the label \Z boson is defined to also include contributions from virtual photons and the interference between the \Z boson and the virtual photon.\label{note}} 
%provide important tests of 
is highly interesting for probing
%tests of
the quantum chromodynamics (QCD) and electroweak (EW) sectors.
%of the Standard Model. 
Particularly, precision measurements of the \Z boson production cross-section at various experiments offer valuable insights for 
testing 
%theoretical 
Standard Model
predictions~\cite{Rijken:1994sh,Hamberg:1990np,Harlander:2002wh,vanNeerven:1991gh,Anastasiou:2003ds,Scimemi:2019cmh,Bury:2022czx,Camarda:2021ict,Duhr:2021vwj}, which are 
%further validated through their comparison 
obtained from precision perturbative QCD calculations 
up to the order of $\alpha_{s}^{3}$~\cite{Camarda:2021ict,Duhr:2021vwj}.

High precision measurements of the \Z boson production cross-sections at different rapidities at \lhcb have imposed important constraints on parton distribution functions (PDFs). 
Results from deep inelastic scattering and hadronic collisions~\cite{H1:2009pze,H1:2015ubc,CDF:1996zyp,CDF:2005cgc,CDF:2010vek,CDF:2008hmn,D0:2007pcy,D0:2006eri,D0:2014kma,D0:2008nou,ATLAS:2011qdp,ATLAS:2015iiu,CMS:2013pzl,CMS:2012ivw,ATLAS:2019zci,CMS:2019raw}, parameterized in terms of the Bjorken variable, $x$, indicating the fraction of the proton momentum carried by a single
parton, are used in the global fits of the PDFs. However, these measurements provide limited information on the PDFs in the
very large ($x$ $\sim0.8$) or very small ($x$ $\sim10^{-4}$) Bjorken-$x$ regions. This leads to large
uncertainties on the PDFs, and on the theoretical predictions that make use of
them. Forward acceptance of the \lhcb detector covers a unique region of phase space, allowing 
measurements of highly boosted \Z boson candidates to be made. Measurements within this
region are sensitive to both large and small Bjorken-$x$ values.
Previous measurements of single \W and \Z production by the \lhcb collaboration~\cite{LHCb-PAPER-2014-033,LHCb-PAPER-2015-001,LHCb-PAPER-2015-003,LHCb-PAPER-2015-049,LHCb-PAPER-2016-024,LHCb-PAPER-2016-021,LHCb-PAPER-2021-037} 
have been included in PDF determinations~\cite{Dulat:2015mca,Harland-Lang:2014zoa,NNPDF:2014otw,Hou:2019efy,NNPDF:2017mvq,NNPDF:2021njg,Bailey:2020ooq,Eskola:2009uj,Martin:2009iq}
and 
%; the \lhcb data contribute 
significantly contributed to the precision of the quark PDFs at large and small values of $x$.

In addition, the \Z boson production cross-section is useful for 
constraining nuclear PDFs (nPDFs), 
providing a clean probe of nuclear-matter effects in the initial state. These effects are typically studied in terms of the nuclear modification factor, 
\rpa, defined as the ratio of the yield observed in \pPb collisions to that in $pp$ collisions, scaled by the mean number of nucleon-nucleon interactions. 
This quantity is used to study the modification of particle production in heavy-ion collisions compared to $pp$ collisions. The \lhcb experiment published the first inclusive \Z production result 
in \pPb collisions at a nucleon-nucleon centre-of-mass energy of $\sqrt{s_{NN}}=5.02\tev$~\cite{LHCb-PAPER-2014-022}.
However, no nuclear modification factor has been reported so far for this collision energy due to the absence of a cross-section measurement in $pp$ collisions. 
With this new measurement of the \Z boson production cross-section in $pp$ collisions at $\sqs=5.02\tev$, 
the nuclear modification factors in the forward region and backward regions are reported here for the first time. 
The forward and backward regions, defined by the muon rapidity $y^*_{\mu}$ in the \pPb centre-of-mass frame, are $1.53<\murapstar<4.03$ and $-4.97<\murapstar<-2.47$, as the \pPb collision system experiences an asymmetric distribution of beam energy.

In this article, the integrated and differential \Z boson production cross-sections in different kinematic bins are measured at the Born level in QED,
using $pp$ collision data collected by the \lhcb detector at a centre-of-mass energy of $\sqs = 5.02\tev$ in 2017, corresponding to an integrated luminosity of $100\invpb$~\cite{LHCb-PAPER-2014-047}. 
The production cross-sections are measured in a fiducial region that closely matches the acceptance of the \lhcb detector, following the analysis strategy developed in Ref.~\cite{LHCb-PAPER-2021-037}.
The fiducial region is defined by requiring that both muons have a pseudorapidity in the range of $2.0<\eta<4.5$ and transverse momentum $\pt>20\gevc$, and that the
di-muon invariant mass is in the interval $60<M_{\mu\mu}<120\gevcc$.

%%%%%%%%%%%%%%%%%%%%%%%%%%%%%%%%%%%%%%%%%%%%%%%%%%%%%%%%%%%%%%%%%%
%%% detector
\section{Detector and simulation}
\label{sec:Detector}
The \lhcb detector~\cite{LHCb-DP-2008-001,LHCb-DP-2014-002} is a single-arm forward
spectrometer covering the \mbox{pseudorapidity} range $2<\eta <5$,
designed for the study of hadrons containing \bquark or \cquark
quarks. The detector includes a high-precision tracking system
consisting of a silicon-strip vertex detector surrounding the $pp$
interaction region~\cite{LHCb-DP-2014-001}, a large-area silicon-strip detector (TT)~\cite{LHCb-TDR-009}, located
upstream of a dipole magnet with a bending power of about
$4{\mathrm{\,Tm}}$, and three stations of silicon-strip detectors and straw
drift tubes~\cite{LHCb-DP-2017-001} placed downstream of the magnet.
The tracking system provides a measurement of the momentum, \ptot, of charged particles with
a relative resolution that varies from 0.5\,\% at low momentum to 1.0\,\% at 200\gevc.
The minimum distance of a track to a primary $pp$ collision vertex, the impact parameter, 
is measured with a resolution of $(15+29/\pt)\mum$,
where \pt is the component of the momentum transverse to the beam, in\,\gevc.
Photons, electrons and hadrons are identified by a calorimeter system consisting of
scintillating-pad and preshower detectors, an electromagnetic
% calorimeter
and a hadronic calorimeter. 
Muons are identified by a system composed of alternating layers of iron and multiwire
proportional chambers~\cite{LHCb-DP-2012-002}.
The online event selection is performed by a trigger~\cite{LHCb-DP-2012-004}, 
which consists of a hardware stage, based on information from the calorimeter and muon
systems, followed by a software stage, which applies a full event reconstruction.

Simulation is used to model the effects of the detector acceptance and the imposed selection requirements.
In the simulation, $pp$ collisions are generated using \pythia~\cite{Sjostrand:2006za,Bierlich:2022pfr} 
with a specific \lhcb configuration~\cite{LHCb-PROC-2010-056}.
Final state radiation (FSR) is generated using \photos~\cite{davidson2015photos}. 
The interactions of the generated particles with the detector, and its response,
are implemented using the \geant toolkit~\cite{Allison:2006ve} as described in Ref.~\cite{LHCb-PROC-2011-006}. 

In this paper, three generators are used to calculate the theoretical predictions.
The \resbos~\cite{Balazs:1997xd} program performs resummation of large logarithms, achieving accuracy up to the next-to-next-to-leading-logarithm level,
%up to next-to-next-leading-logarithm accuracy, 
within the Collins-Soper-Sterman resummation formalism~\cite{Collins:1984kg,Collins:1981uk,Collins:1981va} and matches to next-to-leading-order (NLO) fixed order calculations. 
It also provides theoretical predictions for processes involving FSR corrections and event generation.
The enhanced QCD prediction is obtained from the simulated \Zmm sample, generated with \resbos~\cite{Balazs:1997xd} using the CT18 PDFs~\cite{Hou:2019efy} for all measurements. 
\powheg-BOX~\cite{Nason:2004rx,Frixione:2007vw,Alioli:2008gx,Alioli:2010xd} is an NLO generator, 
and can be interfaced with \pythia for QCD and EW showering. 
Although \pythia is a leading-order (LO) generator, it approximates higher order effects in initial and final states via a parton showering algorithm~\cite{Sjostrand:2007gs}.
In this analysis, \powheg-BOX is used to generate \Zmm events, followed by hadronization using \pythia for all measurements. 
The {\sc MCFM} package~\cite{Campbell:2019dru}, a fixed-order next-to-next-to-leading-order (NNLO) generator, is
used here to estimate the \Z boson production cross-section as a function of \zy in the acceptance of the \lhcb detector. 
For these three generators, different PDFs sets NNPDF3.1~\cite{NNPDF:2017mvq}, NNPDF4.0~\cite{NNPDF:2021njg}, MSHT20~\cite{Bailey:2020ooq}, 
and CT18~\cite{Hou:2019efy} are employed to provide theoretical predictions. 

%%%%%%%%%%%%%%%%%%%%%%%%%%%%%%%%%%%%%%%%%%%%%%%%%%%%%%%%%%%%%%%%%%
%%% event selection
\section{Event selection and background estimation}
\label{sec:sel}
The muon triggers are responsible for the online event selection. At the hardware trigger stage, a muon candidate with high \pt is required. 
The muon candidate is required to have $\pt > 6\gevc$ and $p > 8\gevc$, along with a good track fit quality in the first software trigger stage. 
In the second software trigger stage, an additional requirement of $\pt > 12.5\gevc$ is imposed on the muon candidate. 
For a \Zmm candidate, it is necessary for at least one of the muons to pass both the hardware and software trigger stages. 

\begin{figure}[tb]
\begin{center}
  \includegraphics[width=0.7\textwidth]{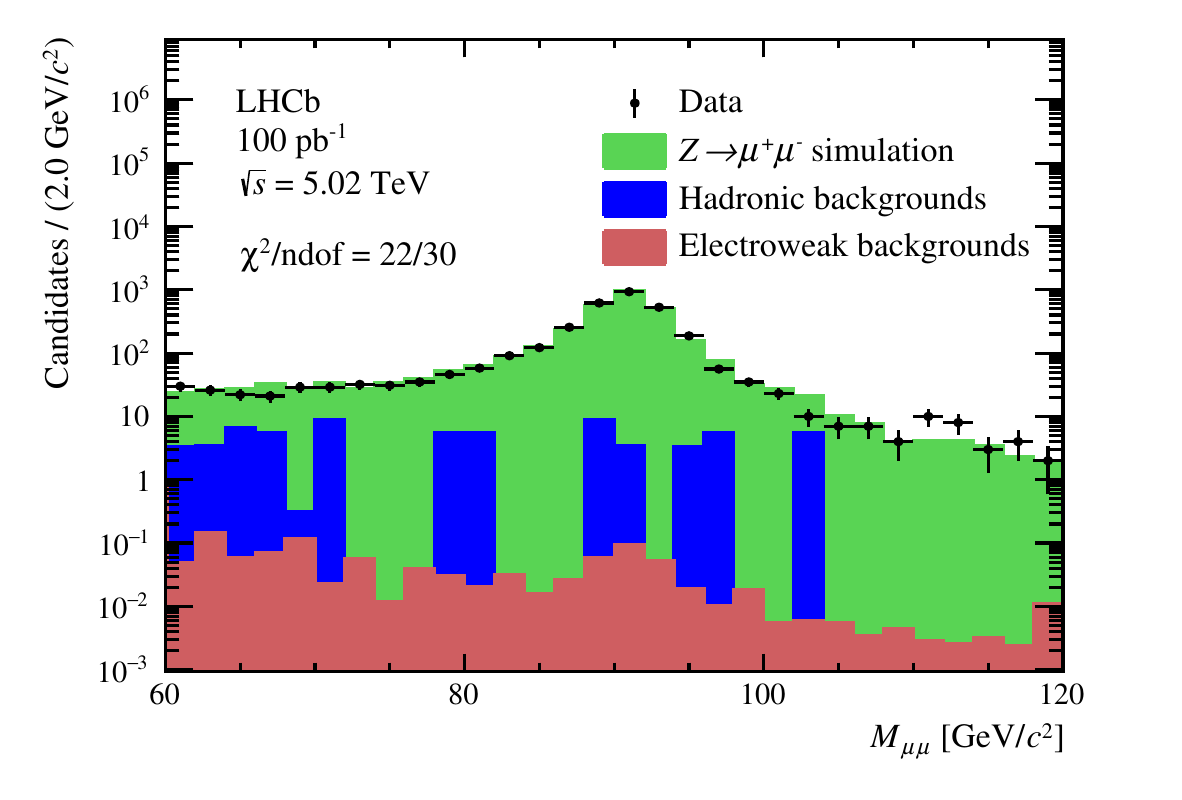}
\caption{Mass distribution of the \Zmm signal candidates. The data are overlaid with model of the signal and background models. 
The signal component is scaled such that the sum of the signal and background matches the integral of the data.}
\label{fig:mass}
\end{center}
\end{figure}

A high-purity \Zmm sample is reconstructed from a pair of opposite-signed tracks identified as muons. 
The invariant mass of the di-muon is required to be within the range $60 <M_{\mu\mu}< 120 \gevcc$. For each muon track, 
the fiducial requirements are $\pt>20\gevc$ and pseudorapidity in the range $2.0<\eta<4.5$. The muons are required to have momentum measurements with relative uncertainties below 10\%. 
In total, 3265 \Zmm candidates meet these selection criteria, and the distribution of the di-muon invariant mass for the selected candidates is shown in Fig.~\ref{fig:mass}.

The background contribution from decays of heavy-flavour hadrons is estimated using two control samples. These samples are used as two independent background determinations, which allow for cross-checking with each other.
Tracks from heavy-flavour decays degrade the primary vertex (PV) fit quality when included in the PV fit, 
since heavy-flavour hadrons travel a finite distance before decaying. 
Hence, the first control sample is obtained by applying a requirement that the selected candidate have a PV with a low fit quality ($\chi^2>95$). 
Additionally, muons produced from semileptonic decays of heavy flavour hadrons are less isolated. The variable $I_{\mu}$ is defined as the ratio of the muon \pt to the vector sum of the \pt of all charged particles 
in a cone of size $R=\sqrt{(\Delta\eta)^{2}+(\Delta\phi)^{2}} < 0.5$ around the muon.
The second control sample is selected by requiring that the two muons are not spatially isolated ($I_{\mu}<0.91$) from the rest of the event.

The yields of these two control samples are determined from a fit to the di-muon invariant mass distributions, using an exponential function. From the fitting results in Appendix~\ref{app:hf}, it can be observed that statistical uncertainty is the primary source of uncertainty in this estimation. 
The heavy-flavour background is concentrated at low mass (${50<M_{\mu\mu}<70\gevcc}$), so to obtain a larger sample and a more stable fit, 
the background yield is determined in the region ${50<M_{\mu\mu}<110\gevcc}$.
% The estimated yield of background is corrected for the efficiency of vertex quality or isolation selection and then extrapolated to the signal region ($60<M_{\mu\mu}<120\gevcc$).
The estimated background yield is corrected for the efficiency of vertex and isolation selections, and extrapolated to the signal region ($60<M_{\mu\mu}<120\gevcc$).
%The extrapolation corrects for the efficiency of the vertex quality and isolation requirements.
The efficiency of the muon isolation (vertex quality) selection is calculated separately by applying the muon isolation requirement (vertex quality requirement) to the first (second) control sample.
%The estimated background contributions, estimated using these two independent samples are consistent, 
The background contributions estimated using these two independent samples are consistent,
and the average value 
is taken as the background contribution from the heavy flavour decay process,
which is determined to be $(2.0\pm1.6) \times 10^{-2}$ for the selected \Zmm sample.

The contribution from the combinatorial background including misidentified hadrons and $B$-$\bar{B}$ mixing is estimated using pairs of same-sign muons in the data.
In the same-sign events, a muon from heavy flavour decay combined with a misidentified hadron are expected to make sizable contributions. 
After removing the contribution from heavy flavour processes, the contribution from misidentified hadrons is determined to be $(6.0\,\pm\,3.0) \times 10^{-5}$, which is negligible.
The electroweak background contributions from the $\ttbar$, $\ww$, $\wz$, $\zz$ and \Ztt processes are estimated to be $(4.4\,\pm\,2.9)\times 10^{-5},\, (5.8\,\pm\,0.7)\times 10^{-5},\, (1.8\,\pm\,0.5)\times 10^{-5},\, (5.9\,\pm\,1.2)\times 10^{-5}$ and $(1.2\,\pm\,0.4)\times 10^{-4}$ from the simulation, 
with LO to NNLO correction factors determined with the 
{\sc MCFM}~\cite{Campbell:2019dru} package.

In summary, the total background contribution to the \Zmm sample in the mass range $60 < M_{\mu\mu} < 120 \gevcc$ is determined to be $(2.0\pm1.6) \times 10^{-2}$ and the background composition of the candidate sample is 
summarised in Table~\ref{tab:totalbkg}. 
%The data-driven backgrounds refer to processes involving heavy flavor decays and hadron misidentification, which are estimated based on the observed events in the data.

\begin{table}[!htbp]
\begin{center}
\caption{Summary of the background composition in the \Zmm data sample of candidates satisfying the signal selection. 
}
\begin{tabular}{lccc}
      Background  & Estimation  &  Events & Fraction\\
\hline
\rule{-4pt}{12pt}
      Heavy flavour (\bbbar, \ccbar)  &    data-driven  &  $65 \pm 51 $  & $2.0 \times 10^{-2}$ \\ 
      Hadron misidentification      &    data-driven  &  $0.20 \pm 0.10 $ & $6.0 \times 10^{-5}$\\
      \Ztt   &    Simulation         &  $0.39 \pm 0.12$  & $1.2 \times 10^{-4}$ \\
      $WZ/ZZ$ & Simulation           &  $0.25 \pm 0.04 $ & $7.7 \times 10^{-5}$ \\
      $WW$ & Simulation               &  $0.19 \pm 0.02 $ & $5.8 \times 10^{-5}$ \\
      \ttbar &    Simulation         &  $0.14 \pm 0.10$  & $4.4 \times 10^{-5}$\\
\hline
       Total               &                          &  $66 \pm 51 $   & $2.0 \times 10^{-2}$\\
\end{tabular}
\label{tab:totalbkg}
\end{center}
\end{table}

%%%%%%%%%%%%%%%%%%%%%%%%%%%%%%%%%%%%%%%%%%%%%%%%%%%%%%%%%%%%%%%%%%
%%% Cross-section determination
\section{Cross-section determination}
Only single-differential cross-section measurements are performed due to the limited sample yields. The differential cross-section is measured as a function of \zy, \zpt and \phistar~\cite{Banfi:2010cf}, which is defined as
\begin{equation}
    \phistar = \tan\left[(\pi-\Delta\phi^{\ell\ell})/2\right]\sin(\theta^*_{\eta}),
\end{equation}
where $\Delta\phi^{\ell\ell}$ represents the difference in the azimuthal angle between the two muons in the laboratory frame. 
The variable $\theta^*_{\eta}$ is defined by $\cos(\theta^*_{\eta}) = \tanh [(\eta^{-} - \eta^{+}) /2]$, with $\eta^-$ and $\eta^+$ denoting the pseudorapidities of the negatively and positively charged muons in the laboratory frame. 
The observable $\phistar$ probes similar physics as the transverse momentum \zpt, but is measured with near-perfect resolution.

The integrated cross-section is obtained by integrating over all bins, which are chosen based on the detector resolution and sample size. 
The differential cross section in a generic variable $a$ is defined as
\begin{equation}
\frac{d\sigma_{\Zmm}}{da}(i)= \frac{N_{Z}(i)\cdot f_{FSR}^{Z}(i)}{\lum \cdot \varepsilon^{Z}(i) \cdot \Delta a(i)},
\label{eq:cs}
\end{equation}
where the generic variable $a$ represents the observable \zy, \zpt or \phistar, the index $i$ indicates the bin of the variable under study,
$N_{Z}(i)$ is the signal yield in bin $i$ after background
subtraction, $f_{FSR}^{Z}(i)$ is the FSR correction factor (as discussed in Sec.~\ref{sec:fsr}), 
\lum is the integrated luminosity, $\Delta a(i)$ is the bin width for the $i$-th
bin (as presented in Tables~\ref{tab:fsr_1D_y} to~\ref{tab:fsr_1D_phi} in Appendix~\ref{app:fsr}), and $\varepsilon^{Z}(i)$ is the total efficiency in the $i$-th bin.

To account for detector misalignment effects, the \Z mass peak position and resolution in simulated events are corrected to be compatible with the data, 
using momentum scaling and smearing factors\cite{LHCb-PAPER-2021-024}. 
The impact from this correction on the integrated cross-section measurement is found to be negligible.

%%%%%%%%%%%%%%%%%%%%%%%%%%%%%%%%%%%%%%%%%%%%%%%%%%%%%%%%%%%%%%%%%%
%% efficiency
\subsection{Efficiency}
The selection efficiencies are determined for the muon tracking, identification and trigger requirements. These are derived using the \Zmm data and a tag-and-probe method~\cite{LHCb-DP-2013-002}.

In the determination of the tracking efficiency, a particle reconstructed in all the tracking subdetectors, and fulfilling the muon trigger 
and muon identification requirements, is used as the tag. 
An object reconstructed by combining hits in the muon stations and the TT downstream tracking stations, denoted as a MuonTT track, then acts as the probe. 
As described in Ref.~\cite{LHCb-DP-2013-002}, the tracking efficiency is calculated as
the fraction of probe candidates matched with a reconstructed track.
However, the precision of the measured tracking efficiency is limited by the low number of \Z boson candidates in the data sample.
In this analysis, we therefore use the tracking efficiency $\varepsilon^{\rm{MC},5.02}_{\rm{Tracking}}$ found from the 5.02\tev \Zmm simulation, 
and apply a correction to this using a scale factor determined using the tag-and-probe method in the 13\tev analysis~\cite{LHCb-PAPER-2021-037},
as 
\begin{equation}
     \varepsilon^{\rm{Data},5.02}_{\rm{Tracking}} = \varepsilon^{\rm{MC},5.02}_{\rm{Tracking}}\times \frac{\varepsilon^{\rm{\rm{Data},13}}_{\rm{Tracking}}}{\varepsilon^{\rm{MC},13}_{\rm{Tracking}}}.
\label{eq:corr_trk}
 \end{equation}

The muon identification efficiency determination is affected by statistical fluctuations similarly to the tracking efficiency, and the same treatment is applied.

The efficiency of the muon trigger is determined with the tag-and-probe method, where a tag particle is chosen from a particle reconstructed in all tracking subdetectors, 
which must be identified and triggered as a muon. 
The probe particle must pass all selection requirements used in the analysis apart from the trigger requirements. 
The invariant mass of tag and probe particles is further required to be within the range [60,120]\gevcc, 
and the azimuthal separation, $|\Delta\phi|$, greater than 2.7 radians. 
The efficiency is computed as the ratio of the number of probes meeting the muon trigger criteria to the total number of probes. 

After correcting the muon tracking efficiency using Eq.~\ref{eq:corr_trk} and applying this correction method to the muon identification efficiency in a similar way, 
the determined muon tracking reconstruction efficiency $\varepsilon_{\rm{Track}}^{\mu^{\pm}}$, identification efficiency $\varepsilon_{\rm{ID}}^{\mu^{\pm}}$ and trigger efficiency $\varepsilon_{\rm{Trig}}^{\mu^{\pm}}$ vary between
94\% and 98\%, 
90\% and 97\%, 
62\% and 85\%
respectively. 
The efficiencies measured as a function of muon pseudorapidity are presented in Appendix~\ref{app:muon_eff}.

This measurement follows the efficiency correction method employed in the 13\tev analysis~\cite{LHCb-PAPER-2021-037}, which involved a one-dimensional efficiency correction based solely on the muon pseudorapidity. To investigate whether any additional dependencies of efficiencies were overlooked or if the results could be entirely attributed to the 1-D correction, a so-called closure test was performed: the number of reconstructed events in simulation is corrected using the efficiencies determined in simulation, then compared to the true number of events for each of the differential distributions.
%The validity of this correction method is ensured by conducting the so-called closure tests using simulated samples. 
The differences between this correction method and the true information from the simulation is considered as a source of systematic uncertainty (as discussed in Sec.~\ref{sec:syst}).

The \Z event selection efficiency is derived from the muon efficiencies $\varepsilon_{\rm{Track}}$, $\varepsilon_{\rm{ID}}$, $\varepsilon_{\rm{Trig}}$ following
 \begin{equation}
\varepsilon^{Z} =  \left( \varepsilon_{\rm{Track}}^{\mu} \cdot 
\varepsilon_{\rm{Track}}^{\mu} \right) \cdot 
 \left( \varepsilon_{\rm{ID}}^{\mu} \cdot \varepsilon_{\rm{ID}}^{\mu} \right)\cdot 
 \left( \varepsilon_{\rm{Trig}}^{\mu} + 
\varepsilon_{\rm{Trig}}^{\mu} - \varepsilon_{\rm{Trig}}^{\mu} \cdot \varepsilon_{\rm{Trig}}^{\mu}\right).
\end{equation}

\subsection{Bin migration correction}
%% unfolding / Bin migration correction
The detector resolution causes migration between kinematic bins. 
Due to the good angular resolution of the \lhcb detector, negligible migration effects are observed among \zy and \phistar bins. 
% Hence, given the measurements as functions of \zy and \phistar, it is unnecessary to correct the migration effects.
Hence, it is unnecessary to correct for migration effects on the number of events in each bin of \zy and \phistar for the cross-section measurement.

To assess the necessity of correcting the \zpt measurement for bin migrations at the reconstruction level, the ratio of the \zpt distribution before and after applying a Bayesian unfolding procedure~\cite{Prosper:2011zz,DAgostini:1994fjx} is calculated from data. This ratio is then compared to the ratio of the reconstructed to the generated \zpt distribution obtained from the simulated sample. 
As noticeable migration is detected, corrections for migration effects on the \zpt distribution are implemented using the Bayesian unfolding approach mentioned earlier.

%% FSR
\subsection{Final state radiation correction}
\label{sec:fsr}
The measured cross-section is corrected to the Born level in QED,
so that it can be
directly compared with theoretical predictions. 
The final-state radiation (FSR) correction is developed and applied to the measurements, 
by comparing the \resbos~\cite{Balazs:1997xd} predictions with and without the implementation of \photos~\cite{davidson2015photos}, which corrects the quantities of muons after final state radiation to the Born level.
The FSR corrections in bins of \zy, \zpt and \phistar 
are shown in Fig.~\ref{fig:FSR_corr}.
The corrections for single-differential cross-section measurements are also presented in Appendix~\ref{app:fsr}.

\begin{figure}[tb]
\begin{center}
\includegraphics[width=0.48\textwidth]{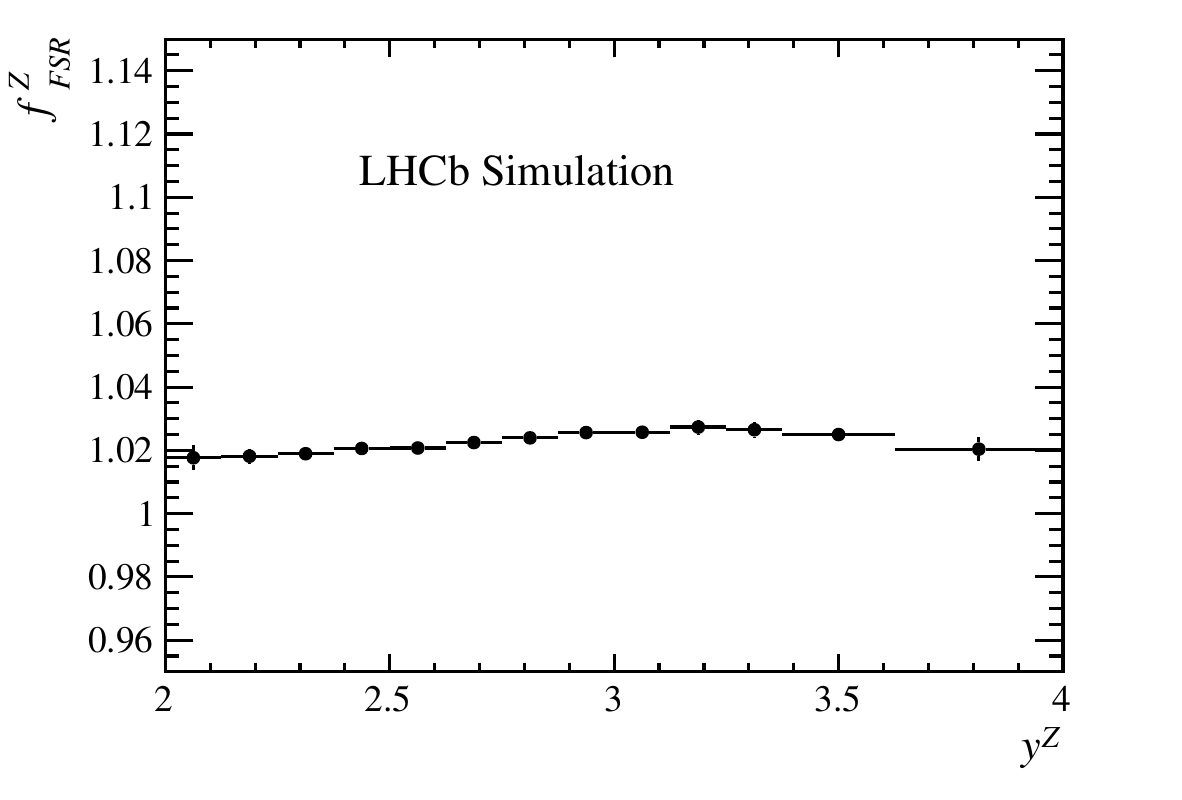}
\includegraphics[width=0.48\textwidth]{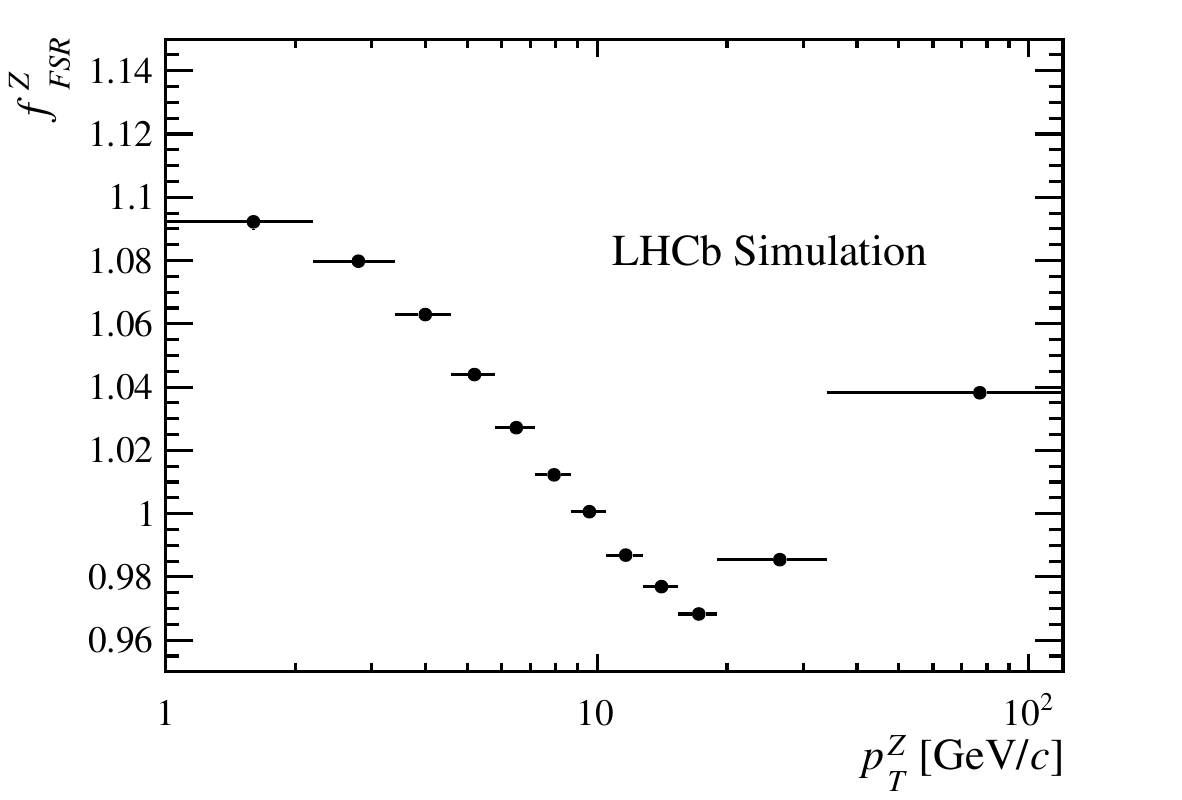}
\includegraphics[width=0.48\textwidth]{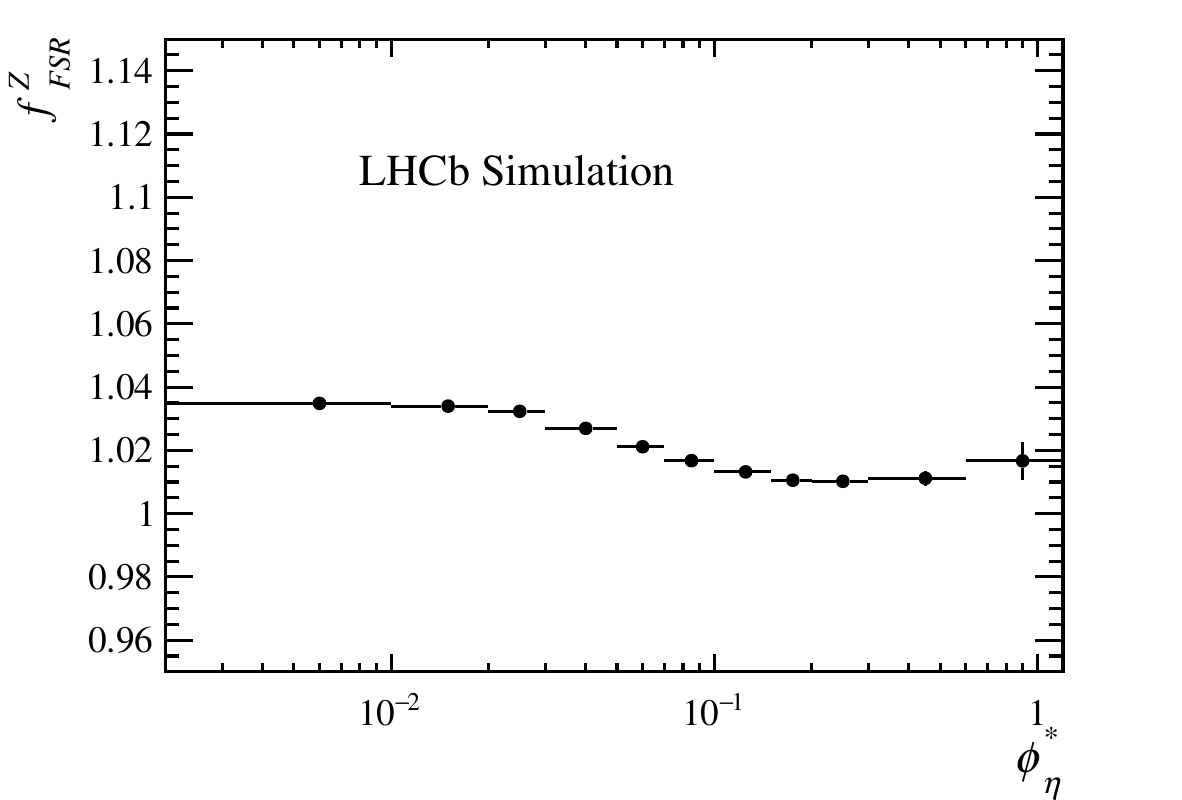}
\caption{Final state radiation correction estimated for the (top-left) \zy, (top-right) \zpt, and (bottom) \phistar differential cross-section measurements.
The error bars represent the total (statistical and systematic) uncertainties.}
\label{fig:FSR_corr}
\end{center}
\end{figure}

%%%%%%%%%%%%%%%%%%%%%%%%%%%%%%%%%%%%%%%%%%%%%%%%%%%%%%%%%%%%%%%%%%
%%%%%%%%  Systematic uncertainties
\section{Systematic uncertainties}
\label{sec:syst}
The systematic uncertainties considered in the present
measurement include the background estimation, the calibration of the momentum scale, the efficiency determination, the bin migration correction, the results of efficiency closure tests, 
the FSR correction, and the measurement of the integrated luminosity.

%% bkg
The determination of the heavy flavour background uses the averaged yield between two methods as the background contribution. 
The associated uncertainty is calculated as the discrepancy between the background yields obtained from the two control samples. 
Additionally, a systematic uncertainty is introduced by varying the mass region and selection requirements of the control samples.
The hadron misidentification and other backgrounds are estimated from the data and simulation, with the statistical uncertainties being treated as systematic uncertainties that depend on the limited size of the same-sign data and simulation samples. 
The systematic uncertainties on the \ttbar, $\ww$, $\wz$, $\zz$, and \Ztt components derive from the statistical uncertainties and theoretical uncertainties on the LO to NNLO correction factors.
To estimate the uncertainty due to inadequate calibration of the detector, the momentum scaling and smearing correction for the simulation are studied in order to improve the modeling of the \Zmm data. 
Only a small fraction, less than 0.01\% of events, exhibits changes when comparing the simulation before and after studying the momentum scaling and smearing. This value is conservatively assigned as the systematic
uncertainty related to the detector alignment. The impact from momentum calibration uncertainty on the integrated cross-section measurement is found to be negligible. However, the alignment uncertainty is considered and added for the \zpt distribution. 
 
%% eff
The track reconstruction and identification efficiencies for high \pt muons are calculated using simulation and data taking at $\sqs = 13\tev$. The systematic uncertainties are determined by the size of the 5.02\tev simulation samples and the uncertainties are propagated from the 13\tev results.
For the trigger efficiency, which is directly measured using control samples in the 5.02\tev data, a systematic uncertainty is assigned for variations due to the limited size of the samples studied. As studied in Ref.~\cite{LHCb-PAPER-2021-037}, an additional systematic uncertainty is evaluated, based on the method used to determine the tracking efficiency. 
This is evaluated to be 0.47\% which is already considered in the systematic uncertainty of the muon tracking efficiency.

%% closure test
To examine whether it is sufficient to perform efficiency corrections only as a function of the muon pseudorapidity variable or if the efficiency also depends on other variables such as muon \pt, a closure test is performed. 
The reconstructed event yields in simulation are adjusted using the efficiencies determined solely based on the muon pseudorapidity obtained from the simulation samples and then compared with the generated yield.
The differences, which do not exhibit any systematic pattern across the differential cross-section measurement regions, are attributed as an additional source of uncertainty.

%% unfold
To estimate the uncertainty attributed to the bin migration correction, the \zpt distribution is unfolded using the so-called {\it{Invert Approach}}~\cite{Adye:2011gm}, which employs a simple inversion of the response matrix without regularisation. 
The deviation of the results from the {\it{Bayesian Approach}}~\cite{Prosper:2011zz,DAgostini:1994fjx} is taken as a systematic uncertainty on the \zpt differential cross-section.
%% FSR
The systematic uncertainty arising from the FSR correction is evaluated by comparing the default correction with one determined using the \powheg generator showered using \pythia. 
The differences in FSR corrections between \resbos with \photos and \powheg with \pythia are then taken into consideration as a systematic uncertainty.
%% lumi
Regarding the data sample used, the luminosity is determined with a precision of 2.0\%~\cite{LHCb-PAPER-2014-047}, which is quoted separately to the other sources of systematic uncertainty. 
The statistical and systematic uncertainties in the integrated cross-section measurement are listed in Table~\ref{tab:uncertainty}. 
The different sources of systematic uncertainty for each bin of the differential cross-sections are summarised in Tables~\ref{tab:err_y} to~\ref{tab:err_phi} in Appendix~\ref{app:results_sys}.

\begin{table}[tb]
\begin{center}
\caption{The uncertainties for the integrated $\Zmm$ cross-section measurement.}
\begin{tabular}{lcc}
\hline
 Source          &  $\Delta \sigma\,$[pb]   & $\Delta\sigma/\sigma$ [\%]   \\\hline
 Luminosity    & 0.79  & 2.00 \\ \hline
Statistical   & 0.70 & 1.77 \\ \hline 
Tracking      & 0.40  & 1.01 \\
Efficiency Closure & 0.24 & 0.61 \\ 
Trigger        & 0.21 &  0.54 \\
Background    & 0.19 & 0.48 \\ 
Identification & 0.10 &  0.25 \\
FSR            & 0.07 & 0.18 \\
Calibration  & $<4.0\times 10^{-3}$ & $<0.01$ \\ \hline 
Total Systematic (excl. lumi.) & 0.56 & 1.42 \\ \hline
\end{tabular}
\label{tab:uncertainty}
\end{center}
\end{table}

%%%%%%%%%%%%%%%%%%%%%%%%%%%%%%%%%%%%%%%%%%%%%%%%%%%%%%%%%%%%%%%%%%
%%% Results
\section{Results}
\subsection{Differential cross-section results}
\begin{figure}[h]
\begin{center}
\includegraphics[width=0.45\textwidth]{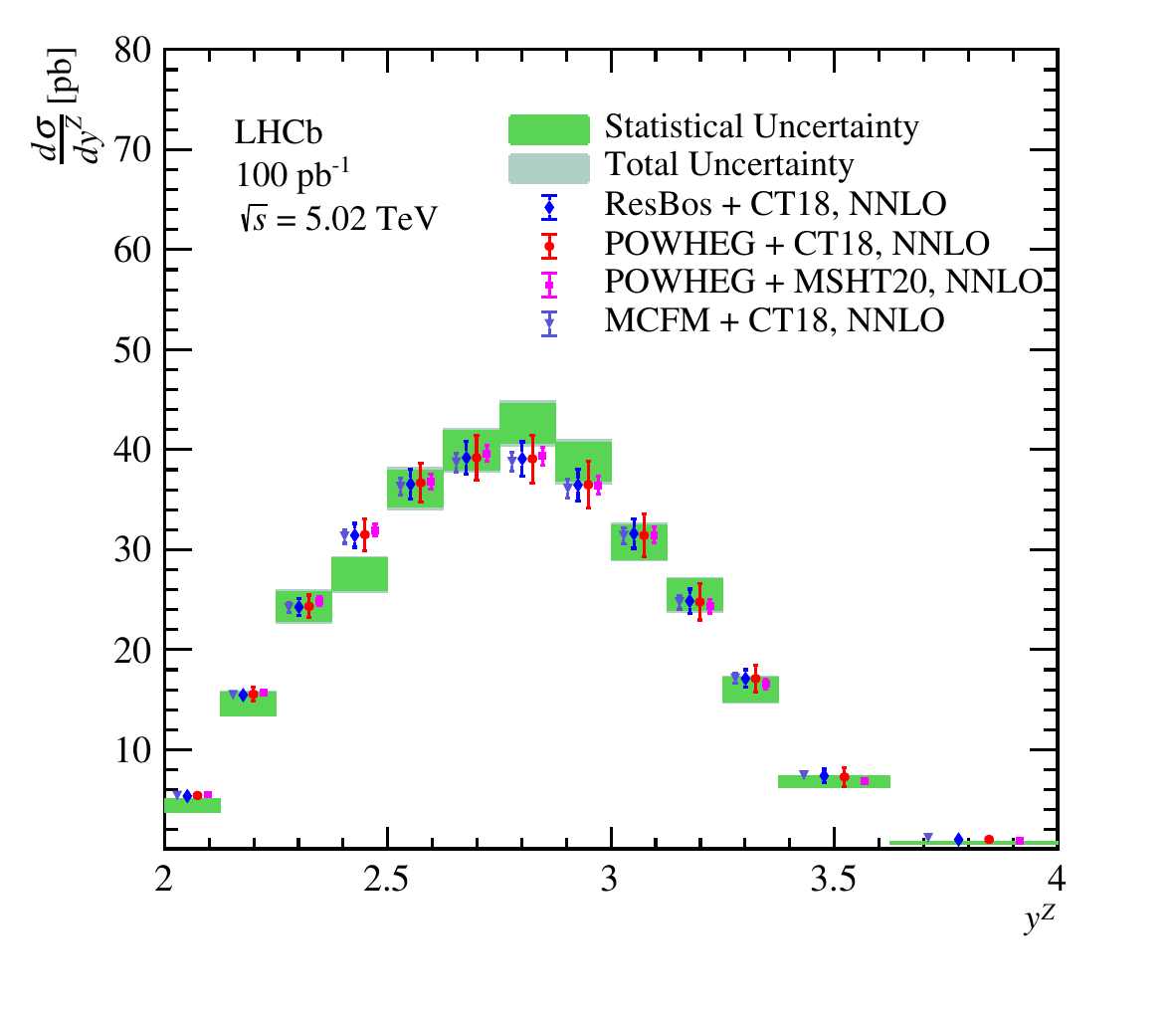}
\includegraphics[width=0.45\textwidth]{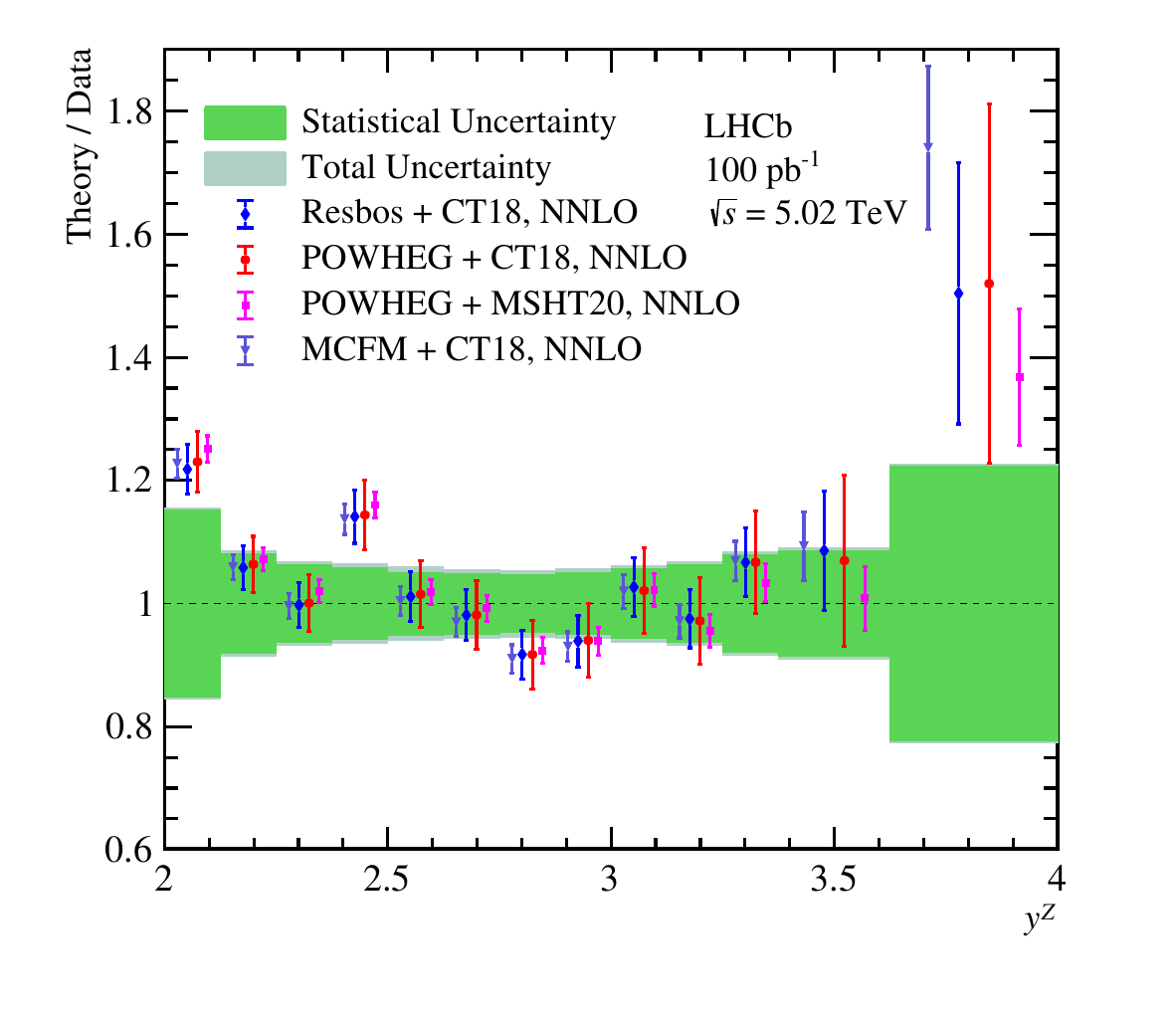}
\includegraphics[width=0.45\textwidth]{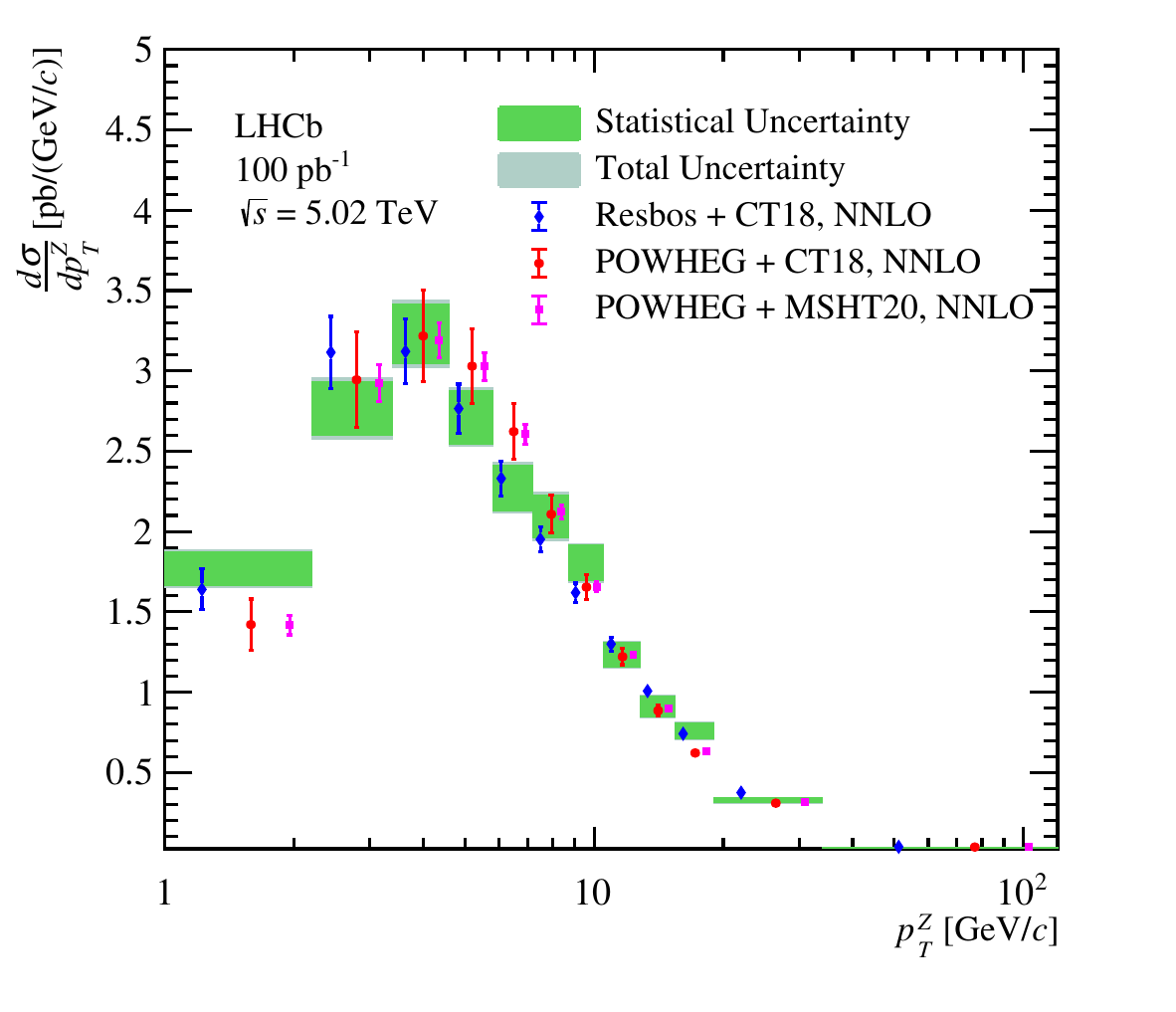}
\includegraphics[width=0.45\textwidth]{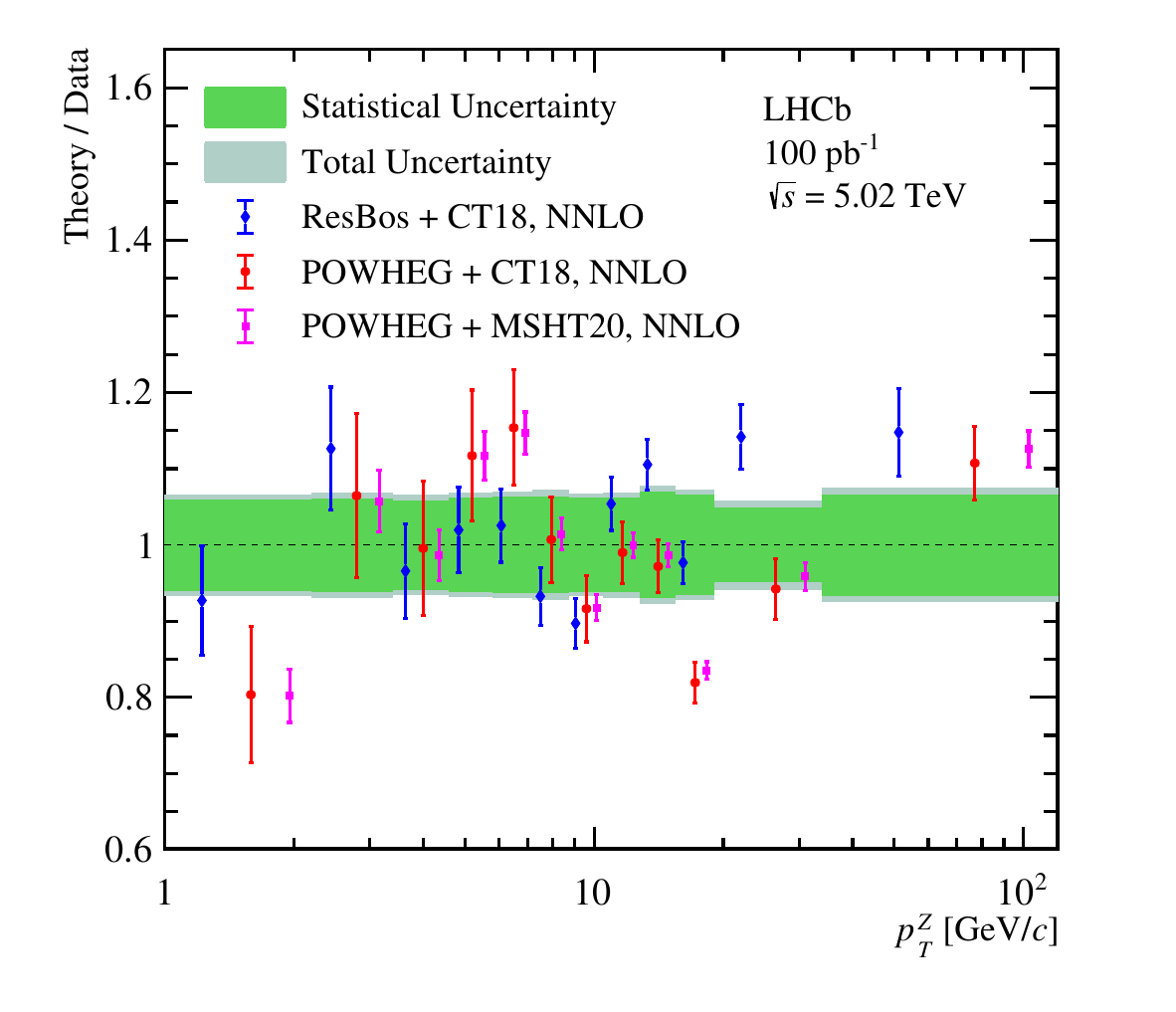}
\includegraphics[width=0.45\textwidth]{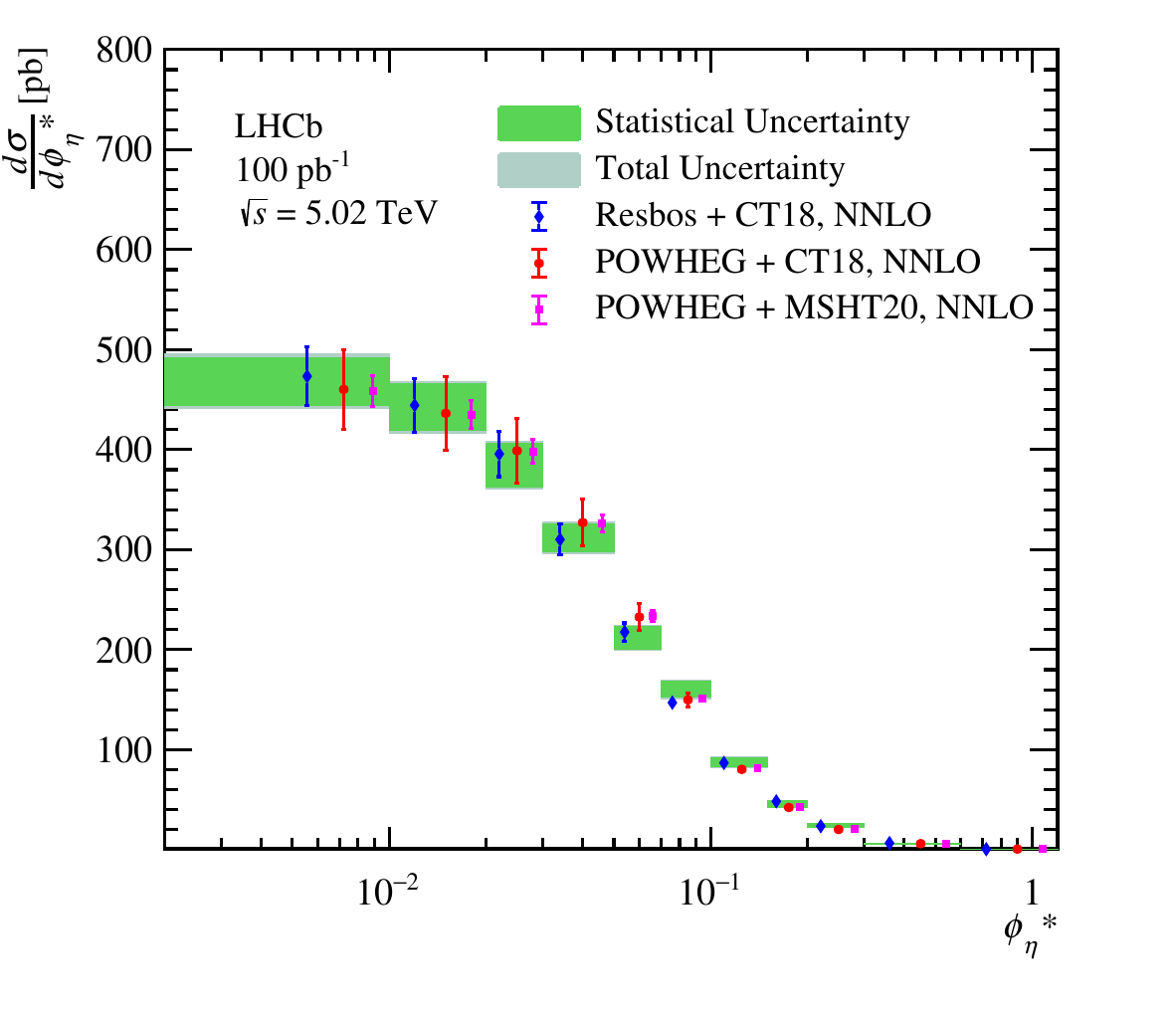}
\includegraphics[width=0.45\textwidth]{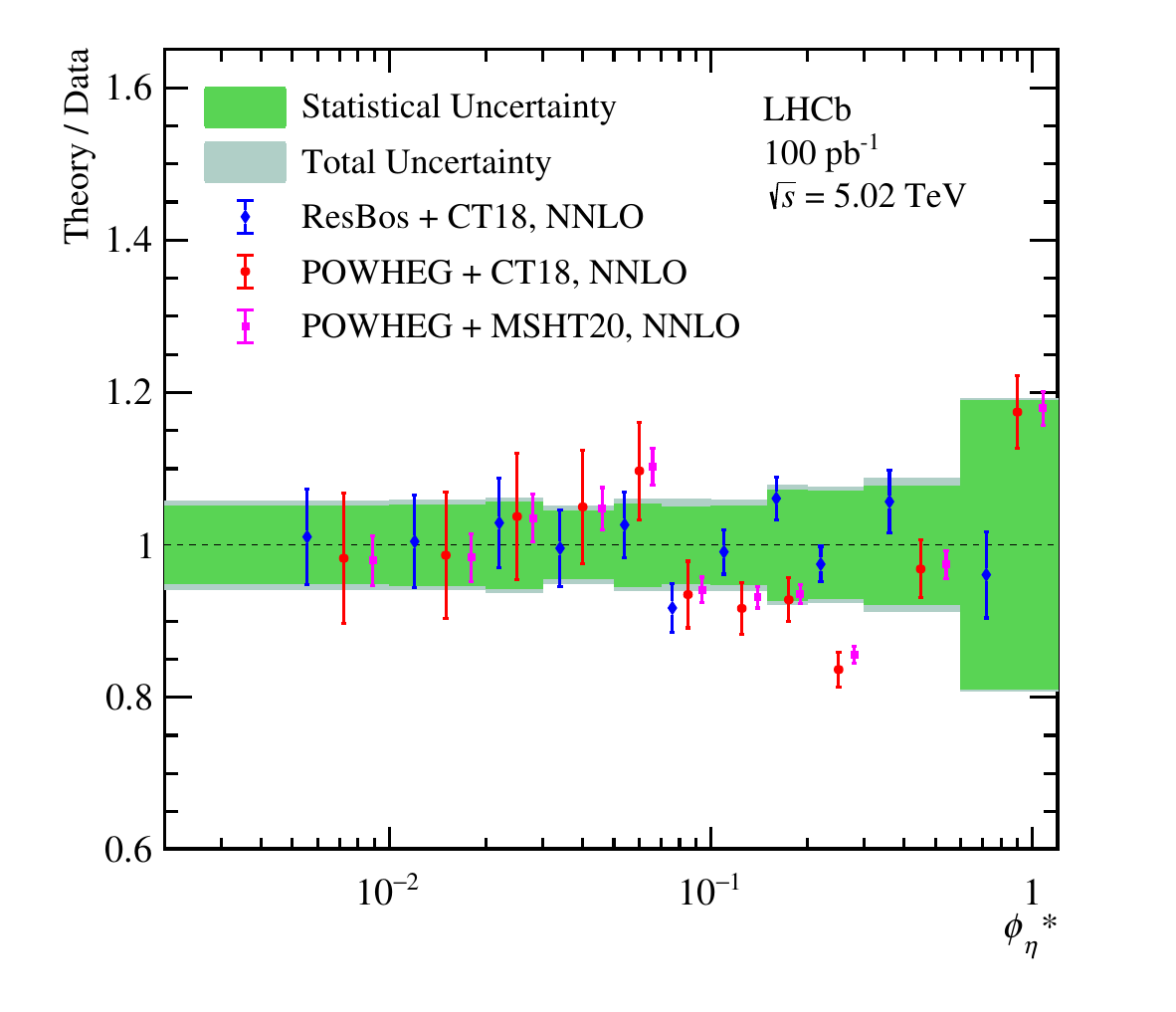}
\caption{(Left) Measured single differential cross-section as a function of \zy, \zpt and \phistar compared with different theoretical predictions. 
(Right) Ratio of theoretical predictions to measured values, with the horizontal bars showing the uncertainty from the PDFs. The green band, centered at unity, shows the uncertainty of the measurement. }
\label{fig:differential_xsec_y}
\end{center}
\end{figure}

The single differential cross-sections in regions of \zy, \zpt and \phistar are shown in Fig.~\ref{fig:differential_xsec_y}, together with the ratios of theoretical predictions to data. The numerical results of single differential cross-section within each bin are summarised in Tables~\ref{tab:cen_y}--\ref{tab:cen_phi} in Appendix~\ref{app:results_cs}.

Measurements are in reasonable agreement with the different theoretical predictions. 
In both the lower and higher \zpt and \phistar region, the measurements agree with predictions from both \resbos and \powheg. For the measurement of \zy, the theoretical predictions from the three generators are also compatible with the experimental results within the uncertainty range.

\subsection{Correlation matrices}
The event migration between bins causes statistical correlations, which are determined using simulation. 
Large correlations are found in the low-\zpt region and small correlations in the high-\zpt region, while the statistical correlations are negligible for both the \zy and \phistar distributions.

For the differential cross-section measurements, background, alignment, efficiency closure test, and FSR uncertainties are assumed to be 50\% correlated between different bins, while the luminosity uncertainty is considered to be 100\% correlated. 
The calculated correlation matrices for the efficiencies are presented in Appendix~\ref{app:correlation}. 
Large correlations between different bins are present in the \zpt differential cross-section measurement, but small correlations are also present between most bins for the \zy and \phistar measurements.

\subsection{Integrated cross-section results}
Using the \lhcb 2017 $pp$ collision data at $\sqs = 5.02\tev$, the integrated Born-level \Z boson production cross-section, with two muons in the final state and within the \lhcb acceptance is

\begin{equation}
\sigma_{\Zmm} = \totxsec
\end{equation}
where the uncertainties are due to statistical effects, systematic effects, and the luminosity measurement, respectively. 

\begin{figure}[tb]
\begin{center}
\includegraphics[width=0.8\textwidth]{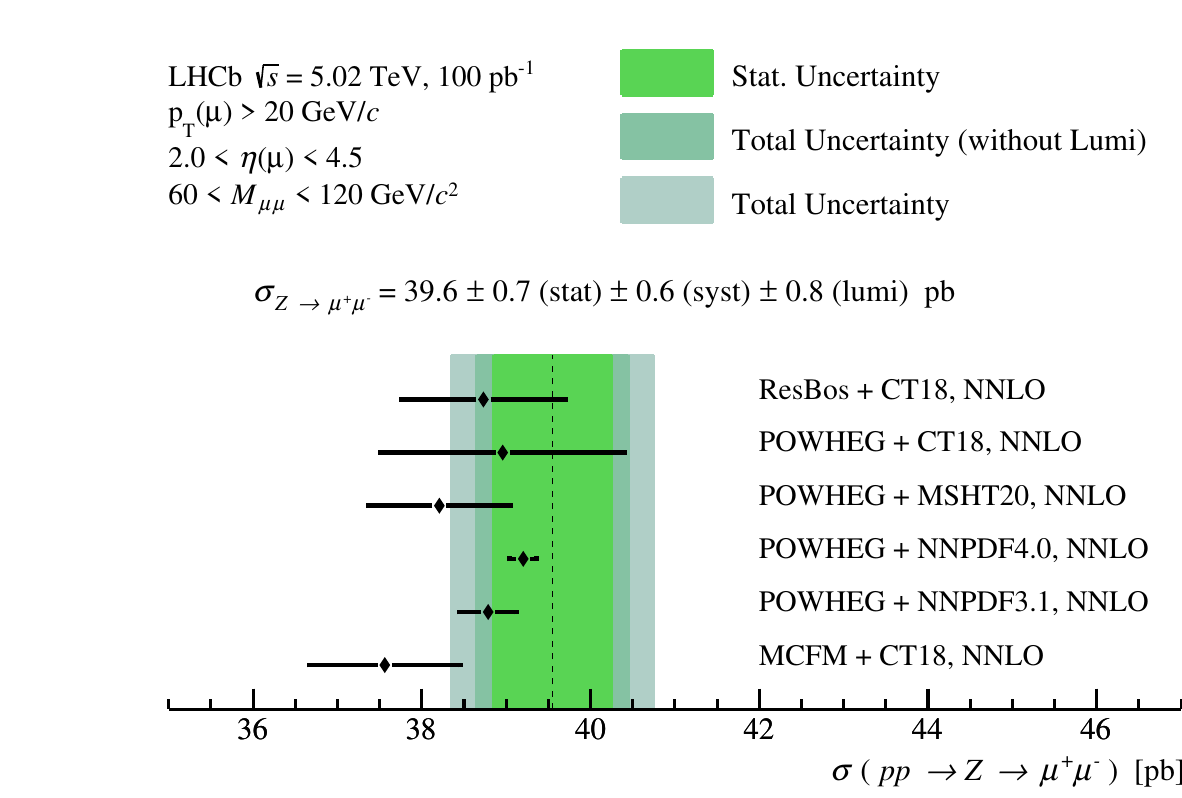}
\caption{Comparison of the integrated cross-section, $\sigma_{\Zmm}$, between data and theoretical predictions. 
The bands correspond to the data, with the inner band corresponding to the statistical uncertainty and the outer bands corresponding to the systematic uncertainty and total uncertainty. }
\label{fig:total_xsec}
\end{center}
\end{figure}

In this article, the $ Z$ boson is defined to also include contributions from virtual photons, and the interference between them since they cannot be distinguished experimentally. 
The measured results obtained for the total cross-section of the $pp \to Z \to \mup \mun$ process at 5.02\tev in the fiducial region of the \lhcb detector 
have been compared to the predictions obtained using {\sc MCFM} with CT18NNLO and other theoretical models, 
including \powheg-Box with NNPDF3.1, NNPDF4.0, MSHT20, CT18, and \resbos with CT18, all of which account for both statistical and PDF uncertainties. 
These theoretical predictions are compared to the results in Fig.~\ref{fig:total_xsec}, which demonstrate a reasonable agreement.

\begin{figure}[!htbp]
\begin{center}
\includegraphics[width=0.8\textwidth]{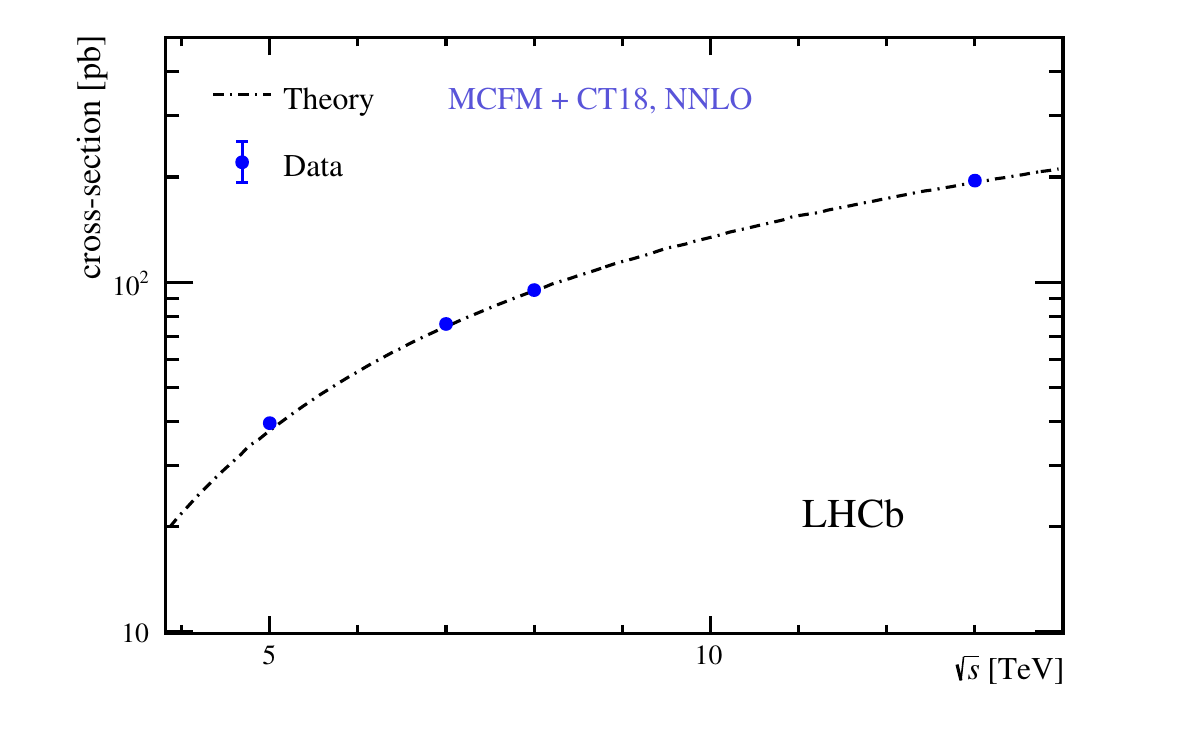}
\caption{Measured $\sigma_{\Zmm}$ for $pp$ collisions, as a function of $\sqrt{s}$. 
The uncertainty on the data is smaller than the
marker size. The data are overlaid with a curve showing the theoretical prediction.
}
\label{fig:mcfm_prediction}
\end{center}
\end{figure}

By comparing the theoretical predictions at different center-of-mass energies provided by {\sc MCFM}, with the experimental measurements previously obtained by \lhcb at $\mbox{\sqs = 7, 8}$ and $13\tev$, as illustrated in Fig.~\ref{fig:mcfm_prediction}, a good level of consistency can be observed overall.


\subsection{Nuclear modification factors}
The nuclear modification factors are determined based on the present measurements and those in Ref.~\cite{LHCb-PAPER-2014-022}.
Here, the statistical and systematic uncertainties (including uncertainties due to integrated luminosity) of the cross-sections measured between $pp$ and \pPb collisions are assumed to be fully uncorrelated when being propagated to the nuclear modification factors. 
Efficiency and background modeling uncertainties are also treated as fully uncorrelated, 
owing to the very different running conditions between $pp$ and \pPb collisions such as the heavier detector contamination from the higher charged hadron multiplicities in case of \pPb. 
All other sources of systematic uncertainty are considered to be uncorrelated, an assumption that has negligible effect to the reported results.

%%%%%%%%%%%%%%%%%%%%%%
The unique forward geometry coverage allows the \lhcb detector to probe
nPDFs at very small Bjorken-$x$ ($10^{-4}<x<10^{-3}$).
The nuclear modification factors can be calculated using the $pp$ cross-section measured here together with the \Z boson production cross-section in \pPb collisions measured in Ref.~\cite{LHCb-PAPER-2014-022}.
The nuclear modification factors are defined as the ratio of the \Z boson production cross-sections between \pPb and $pp$ collisions under the same muon rapidity acceptance.
Due to the asymmetric beam energy in the \pPb centre-of-mass frame, the muon rapidity acceptance ($2.0 < \eta<4.5$) becomes $1.53<\murapstar<4.03$ in case of the forward collisions, and $-4.97<\murapstar<-2.47$ in case of backward collisions,
where $\murapstar$ represents the muon rapidity in the centre-of-mass frame. The two rapidity quantities are identical in $pp$ collisions, $\eta_{\mu}\equiv \murapstar$, due to the symmetric beam energy. Forward collisions refer to the proton beam entering the \lhcb detector along the positive direction of the $z$ axis, while backward collisions correspond to the proton beam going in the opposite direction.

The cross-sections of the \pPb collisions measured under different rapidity acceptances need to be corrected before being used to calculate the nuclear modification factors. 
The \kpa factor corrects for the different $\eta$ acceptance between \pPb and $pp$ collisions
and can be calculated using \powheg with the proton PDF set CTEQ6.1~\cite{Pumplin:2002vw}, 
\begin{equation}
\kpa^{\rm F}= \dfrac{\sigma_{(pp,~2.0<\eta<4.5)}}{\sigma_{(pp,~1.53<\eta<4.03)}} = 0.706\pm 0.002,
\end{equation}
and
\begin{equation}
\kpa^{\rm B}= \dfrac{\sigma_{(pp,~2.0<\eta<4.5)}}{\sigma_{(pp,~-4.97<\eta<-2.47)}} = 1.518 \pm 0.003,
\end{equation}
for forward and backward collisions, respectively. 
The integrated nuclear modification factors can then be defined as follows
\begin{equation}
 \rpa^{\rm F}= \kpa^{\rm F} \cdot \dfrac{\sigma_{(\pPb,~1.53<\murapstar<4.03)}}{ 208 \cdot \sigma_{(pp,~2.0<\eta<4.5)}},
\label{eq:rpa_F_bins}
\end{equation}
and
\begin{equation}
\rpa^{\rm B}= \kpa^{\rm B} \cdot \dfrac{ \sigma_{(\pPb,~-4.97<\murapstar<-2.47)}}{ 208 \cdot \sigma_{(pp,~2.0<\eta<4.5)}},
\label{eq:rpa_B_bins}
\end{equation} 
for forward and backwards \pPb collisions, respectively, with 208 being the number of binary 
nucleon-nucleon collisions in \pPb collisions. 
The quantity $\sigma_{(\pPb,~1.53<\murapstar<4.03)}$ is measured to be \XSecFor, 
and $\sigma_{(\pPb,~-4.97<\murapstar<-2.47)}$ is measured to be \XSecBack~\cite{LHCb-PAPER-2014-022}.

The nuclear modification factors are calculated to be
\begin{equation*}
    R_{pPb}^{\rm F}=\NMFf
\end{equation*}
for the forward collisions, and 
\begin{equation*}
    R_{pPb}^{\rm B}=\NMFb
\end{equation*}
for the backward collisions.
The large statistical uncertainties are due to the small size of the \pPb data sample.

Based on the theoretical prediction of the cross-sections from Ref.~\cite{LHCb-PAPER-2014-022} derived using {\sc FEWZ}~\cite{Li:2012wna} at NNLO with EPS09 nPDFs~\cite{Eskola:2009uj} 
and MSTW08 PDFs~\cite{Martin:2009iq}, the predicted \rpa can be derived as $\rpa^{\rm F, \,\rm theo.} = 0.906^{+0.002}_{-0.007}$ and $R_{pPb}^{\rm B, \,\rm theo.} = 0.929^{+0.011}_{-0.028}$ 
for the forward and backward collisions, respectively. 
The measurement and theoretical prediction agree within the uncertainties in the forward region, 
whereas a 2.86\,$\sigma$ tension between the measured result and the prediction is seen in the backward region.
However, due to the limited size of the \pPb sample, the tension could be caused by a statistical fluctuation.

%%%%%%%%%%%%%%%%%%%%%%%%%%%%%%%%%%%%%%%%%%%%%%%%%%%%%%%%%%%%%%%%%%
%%% Conclusion
\section{Conclusion}
This paper reports the first measurement of \Z boson production cross-section at the centre-of-mass energy $\sqs = 5.02\tev$, using the \lhcb $pp$ collision dataset collected in 2017.
The techniques employed in this analysis closely follow those established in a previous \lhcb Run 2 analysis~\cite{LHCb-PAPER-2021-037}.
The results show reasonable agreement with various predictions in the Standard Model. Combining this measurement with the previous inclusive \Z production result in \pPb collisions at $\sqrt{s_{NN}}=5.02\tev$~\cite{LHCb-PAPER-2014-022}
the nuclear modification factors in the forward and backward regions are obtained for the first time.

%%%%%%% Acknowledgements
\section*{Acknowledgements}
%
% These Acknowledgements valid from 3-May-2019
%
\noindent We express our gratitude to our colleagues in the CERN
accelerator departments for the excellent performance of the LHC. We
thank the technical and administrative staff at the LHCb
institutes.
We acknowledge support from CERN and from the national agencies:
CAPES, CNPq, FAPERJ and FINEP (Brazil); 
MOST and NSFC (China); 
CNRS/IN2P3 (France); 
BMBF, DFG and MPG (Germany); 
INFN (Italy); 
NWO (Netherlands); 
MNiSW and NCN (Poland); 
MEN/IFA (Romania); 
%MSHE (Russia); 
MICINN (Spain); 
SNSF and SER (Switzerland); 
NASU (Ukraine); 
STFC (United Kingdom); 
DOE NP and NSF (USA).
We acknowledge the computing resources that are provided by CERN, IN2P3
(France), KIT and DESY (Germany), INFN (Italy), SURF (Netherlands),
PIC (Spain), GridPP (United Kingdom), 
%RRCKI and Yandex LLC (Russia), 
CSCS (Switzerland), IFIN-HH (Romania), CBPF (Brazil),
Polish WLCG  (Poland) and NERSC (USA).
We are indebted to the communities behind the multiple open-source
software packages on which we depend.
Individual groups or members have received support from
ARC and ARDC (Australia);
Minciencias (Colombia);
AvH Foundation (Germany);
EPLANET, Marie Sk\l{}odowska-Curie Actions, ERC and NextGenerationEU (European Union);
A*MIDEX, ANR, IPhU and Labex P2IO, and R\'{e}gion Auvergne-Rh\^{o}ne-Alpes (France);
Key Research Program of Frontier Sciences of CAS, CAS PIFI, CAS CCEPP, 
Fundamental Research Funds for the Central Universities, 
and Sci. \& Tech. Program of Guangzhou (China);
%Key Research Program of Frontier Sciences of CAS, CAS PIFI,
%Thousand Talents Program, and Sci. \& Tech. Program of Guangzhou (China);
%RFBR, RSF and Yandex LLC (Russia);
GVA, XuntaGal, GENCAT, Inditex, InTalent and Prog.~Atracci\'on Talento, CM (Spain);
SRC (Sweden);
the Leverhulme Trust, the Royal Society
 and UKRI (United Kingdom).

% ===============================================================================
% Purpose: appendix to the standard template: standard symbol alises from Ulrik
% Author: Tomasz Skwarnicki
% Created on: 2009-09-24
% ===============================================================================

%{\noindent\normalfont\bfseries\Large Appendices}
%\newpage

\section*{Appendices}

\appendix

\section{Invariant mass fitting results for background}
\label{app:hf}
\begin{figure}[!htbp]
\begin{center}
\includegraphics[width=0.45\textwidth]{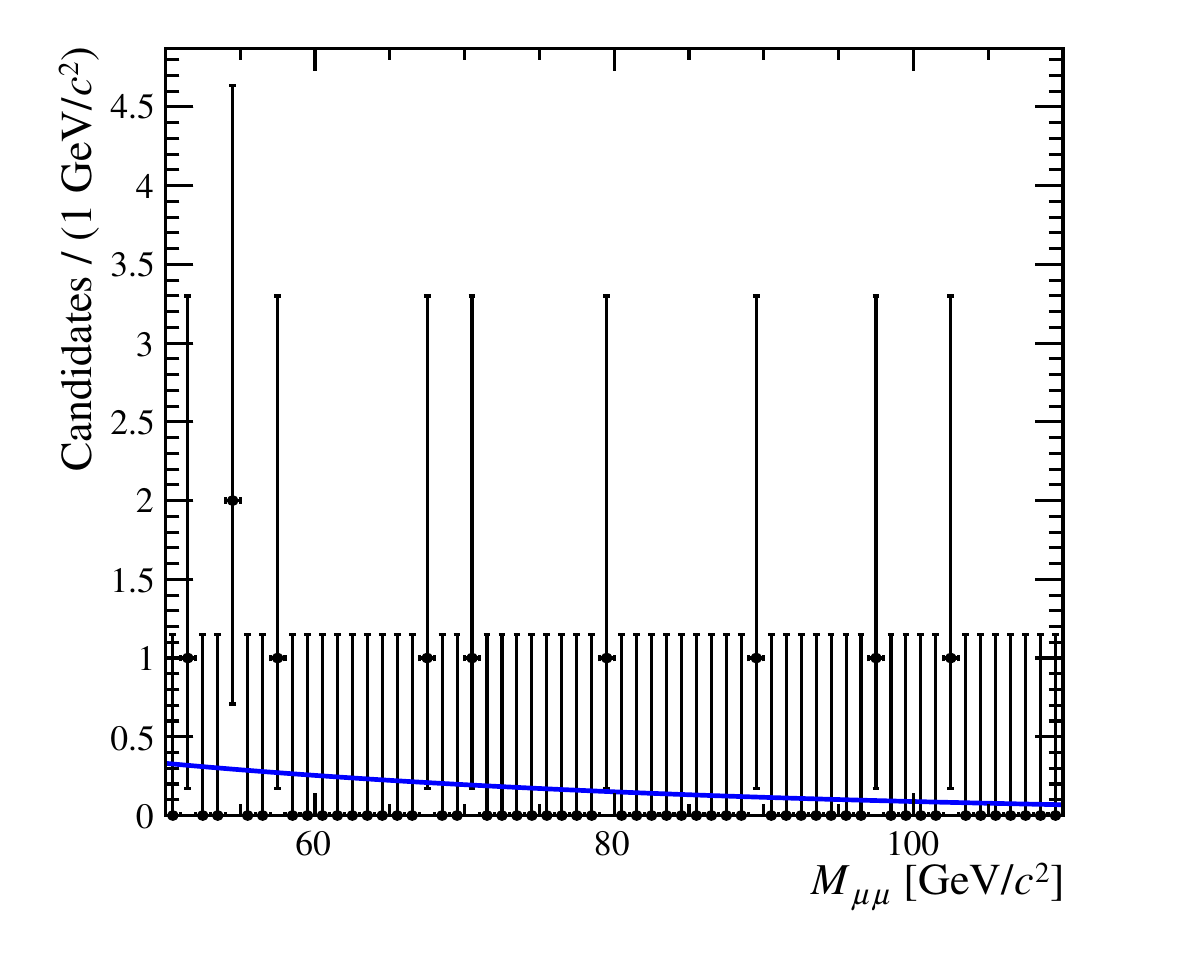}
\includegraphics[width=0.45\textwidth]{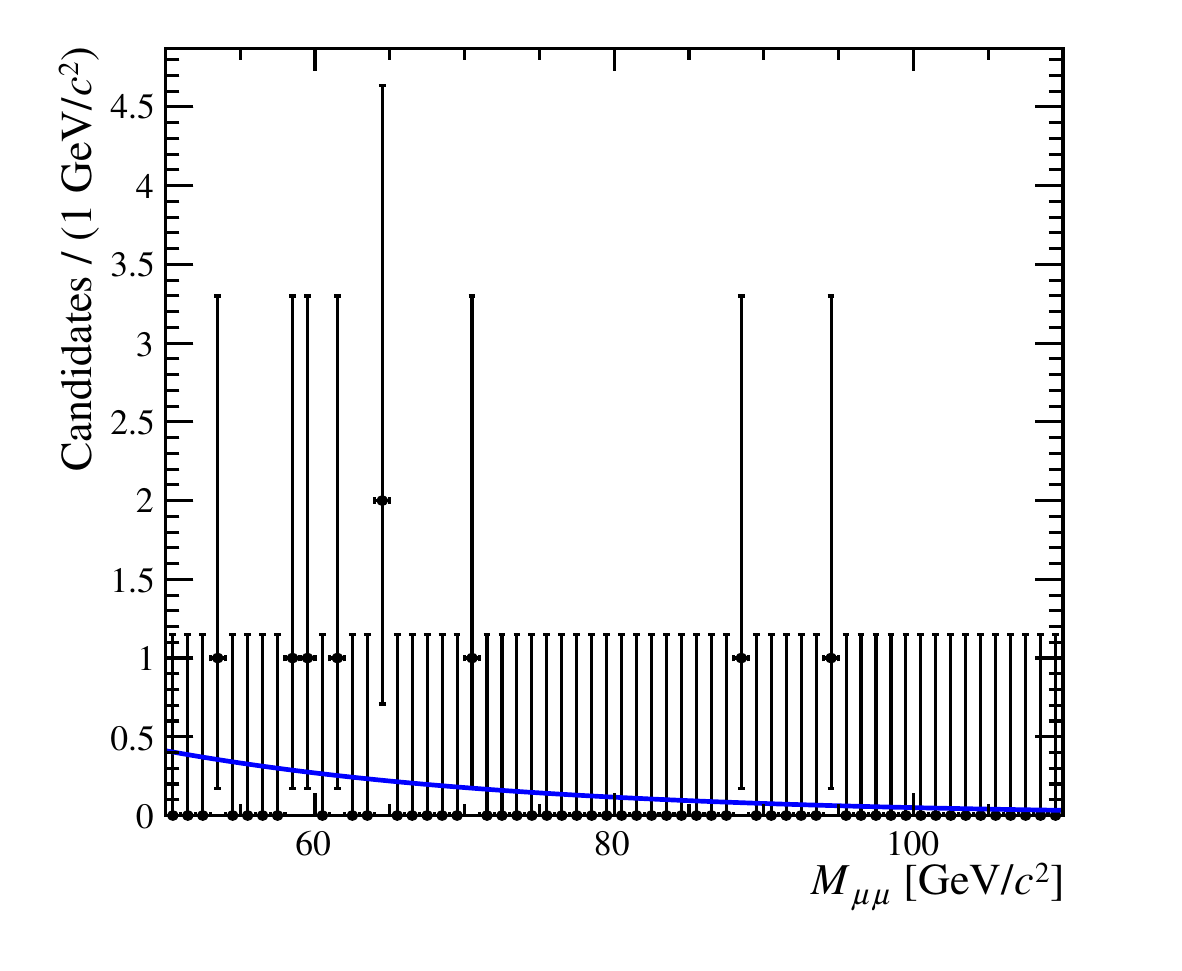}
\caption{Invariant mass distributions of heavy-flavour samples with (left) PV fit quality $\chisq> 95$
and (right) $I_{\mu}< 91\%$ applied to both muons.
Only statistical uncertainties are shown.}
\label{fig:bkgDp}
\end{center}
\end{figure}

To estimate the signal yield for heavy-flavour background-dominated samples, 
an exponential function is used to fit the dimuon invariant mass distributions 
in the region of $50 - 110 \gevcc$, using a binned likelihood method.
The distribution of selected candidates and fitted results are shown in
Fig.~\ref{fig:bkgDp}. The fit result is then 
extrapolated to determine the background contribution in the mass region of $60 - 120 \gevcc$.

\section{Final state radiation corrections}
\label{app:fsr}
Tabulated results of final state radiation corrections used in the single differential cross-section measurements
are presented in Tables~\ref{tab:fsr_1D_y} to~\ref{tab:fsr_1D_phi}.

\begin{table}[tb]
\begin{center}
\caption{Final state radiation correction used in the \zy cross-section measurement. The first uncertainty is statistical and the second is systematic.}
\begin{tabular}{cc}
%% outputs/Systematic/Unc_FSR_ZRapidity_2017.txt
\hline
 \multicolumn{1}{c}{\zy}               & \multicolumn{1}{c}{Correction}  \\ \hline
$[$\,2.000,\, 2.125\,$]$ &  1.018 \,$\pm$\,  0.004  \,$\pm$         0.092 \\ 
$[$\,2.125,\, 2.250\,$]$ &  1.018 \,$\pm$\,  0.002  \,$\pm$         0.000 \\ 
$[$\,2.250,\, 2.375\,$]$ &  1.019 \,$\pm$\,  0.002  \,$\pm$         0.006 \\ 
$[$\,2.375,\, 2.500\,$]$ &  1.021 \,$\pm$\,  0.002  \,$\pm$         0.057 \\ 
$[$\,2.500,\, 2.625\,$]$ &  1.021 \,$\pm$\,  0.001  \,$\pm$         0.046 \\ 
$[$\,2.625,\, 2.750\,$]$ &  1.022 \,$\pm$\,  0.001  \,$\pm$         0.122 \\ 
$[$\,2.750,\, 2.875\,$]$ &  1.024 \,$\pm$\,  0.002  \,$\pm$         0.009 \\ 
$[$\,2.875,\, 3.000\,$]$ &  1.026 \,$\pm$\,  0.002  \,$\pm$         0.162 \\ 
$[$\,3.000,\, 3.125\,$]$ &  1.026 \,$\pm$\,  0.002  \,$\pm$         0.070 \\ 
$[$\,3.125,\, 3.250\,$]$ &  1.027 \,$\pm$\,  0.002  \,$\pm$         0.128 \\ 
$[$\,3.250,\, 3.375\,$]$ &  1.027 \,$\pm$\,  0.002  \,$\pm$         0.011 \\ 
$[$\,3.375,\, 3.625\,$]$ &  1.025 \,$\pm$\,  0.002  \,$\pm$         0.023 \\ 
$[$\,3.625,\, 4.000\,$]$ &  1.020 \,$\pm$\,  0.004  \,$\pm$         0.001 \\
\hline
\end{tabular}
\label{tab:fsr_1D_y}
\end{center}
\end{table}

\begin{table}[tb]
\begin{center}
\caption{Final state radiation correction used in the \zpt cross-section measurement. The first uncertainty is statistical and the second is systematic.}
\begin{tabular}{cc}
%% outputs/Systematic/Unc_FSR_ZPT_2017.txt
\hline
 \multicolumn{1}{c}{\zpt~[\gevc]}              & \multicolumn{1}{c}{Correction}  \\ \hline
$[$\,0.0,\,  2.2$\,]$ &  1.092 \,$\pm$\,  0.002  \,$\pm$  0.020 \\ 
$[$\,2.2,\,  3.4$\,]$ &  1.080 \,$\pm$\,  0.002  \,$\pm$  0.018 \\ 
$[$\,3.4,\,  4.6$\,]$ &  1.063 \,$\pm$\,  0.002  \,$\pm$  0.003 \\ 
$[$\,4.6,\,  5.8$\,]$ &  1.044 \,$\pm$\,  0.002  \,$\pm$  0.015 \\ 
$[$\,5.8,\,  7.2$\,]$ &  1.027 \,$\pm$\,  0.002  \,$\pm$  0.029 \\ 
$[$\,7.2,\,  8.7$\,]$ &  1.012 \,$\pm$\,  0.002  \,$\pm$  0.025 \\ 
$[$\,8.7,\, 10.5$\,]$ &  1.001 \,$\pm$\,  0.002  \,$\pm$  0.003 \\ 
$[$\,10.5,\, 12.8$\,]$ &  0.987 \,$\pm$\,  0.002  \,$\pm$  0.006 \\ 
$[$\,12.8,\, 15.4$\,]$ &  0.977 \,$\pm$\,  0.002  \,$\pm$  0.010 \\ 
$[$\,15.4,\, 19.0$\,]$ &  0.968 \,$\pm$\,  0.002  \,$\pm$  0.005 \\ 
$[$\,19.0,\, 34.0$\,]$ &  0.985 \,$\pm$\,  0.001  \,$\pm$  0.000 \\ 
$[$\,34.0,\, 120.0$\,]$ &  1.038 \,$\pm$\,  0.002  \,$\pm$  0.000 \\
\hline
\end{tabular}
\label{tab:fsr_1D_pt}
\end{center}
\end{table}

\begin{table}[tb]
\begin{center}
\caption{Final state radiation correction used in the \phistar cross-section measurement. The first uncertainty is statistical and the second is systematic.}
\begin{tabular}{cc}
\hline
 \multicolumn{1}{c}{\phistar}              & \multicolumn{1}{c}{Correction}  \\ \hline
$[$\,0.00,\, 0.01\,$]$ &   1.035 \,$\pm$\,  0.002  \,$\pm$  3.802 \\ 
$[$\,0.01,\, 0.02\,$]$ &   1.034 \,$\pm$\,  0.002  \,$\pm$  0.919 \\ 
$[$\,0.02,\, 0.03\,$]$ &   1.032 \,$\pm$\,  0.002  \,$\pm$  1.384 \\ 
$[$\,0.03,\, 0.05\,$]$ &   1.027 \,$\pm$\,  0.001  \,$\pm$  0.878 \\ 
$[$\,0.05,\, 0.07\,$]$ &   1.021 \,$\pm$\,  0.002  \,$\pm$  1.060 \\ 
$[$\,0.07,\, 0.10\,$]$ &   1.017 \,$\pm$\,  0.002  \,$\pm$  0.378 \\ 
$[$\,0.10,\, 0.15\,$]$ &   1.013 \,$\pm$\,  0.002  \,$\pm$  0.220 \\ 
$[$\,0.15,\, 0.20\,$]$ &   1.011 \,$\pm$\,  0.002  \,$\pm$  0.288 \\ 
$[$\,0.20,\, 0.30\,$]$ &   1.010 \,$\pm$\,  0.002  \,$\pm$  0.236 \\ 
$[$\,0.30,\, 0.60\,$]$ &   1.011 \,$\pm$\,  0.002  \,$\pm$  0.039 \\ 
$[$\,0.60,\, 1.20\,$]$ &   1.017 \,$\pm$\,  0.006  \,$\pm$  0.001 \\ 
\hline
\end{tabular}
\label{tab:fsr_1D_phi}
\end{center}
\end{table}

\clearpage

\section{Efficiency}
\label{app:muon_eff}
\begin{figure}[h]
\begin{center}
\includegraphics[width=0.45\textwidth]{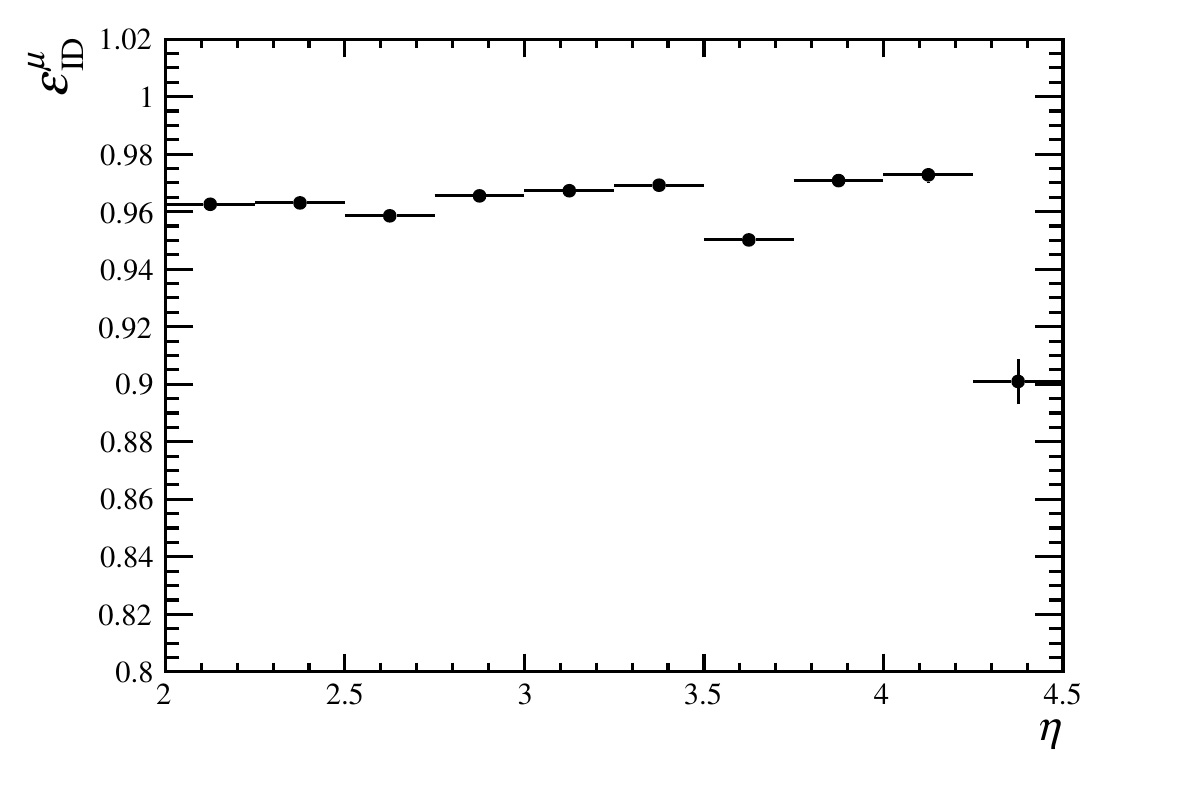}
\includegraphics[width=0.45\textwidth]{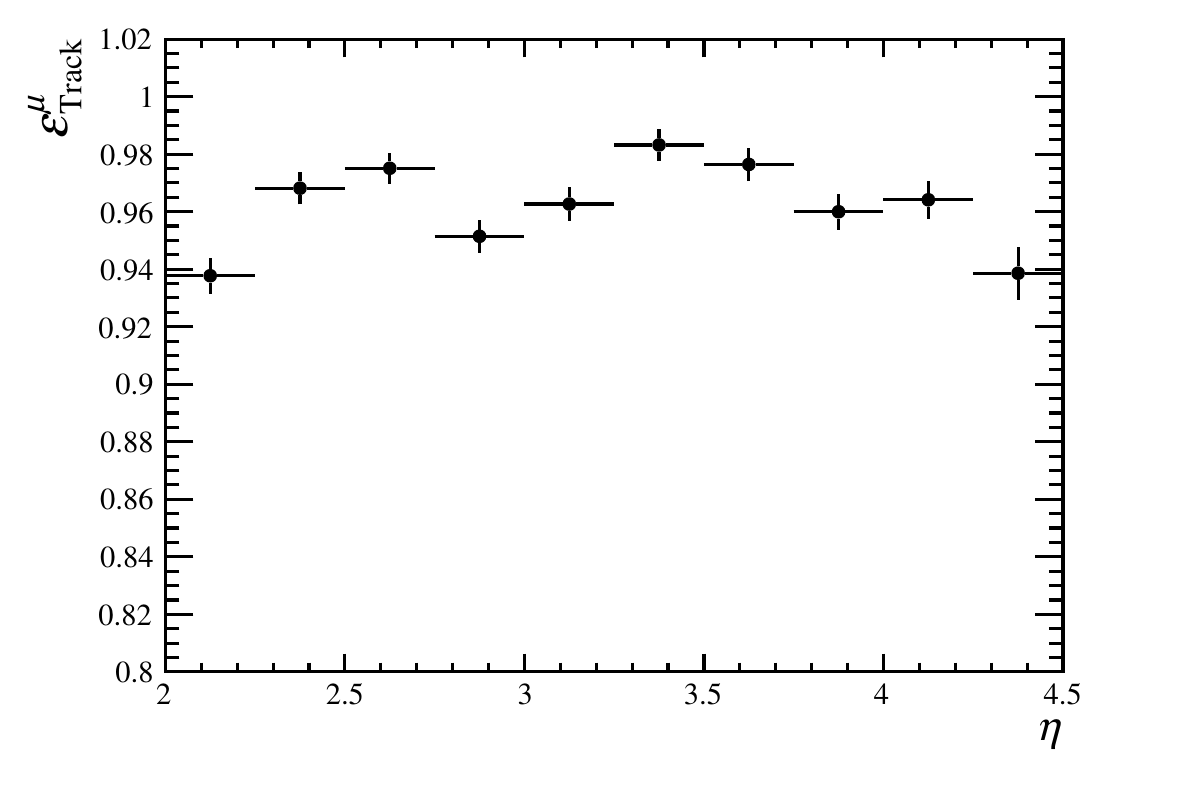}
\includegraphics[width=0.45\textwidth]{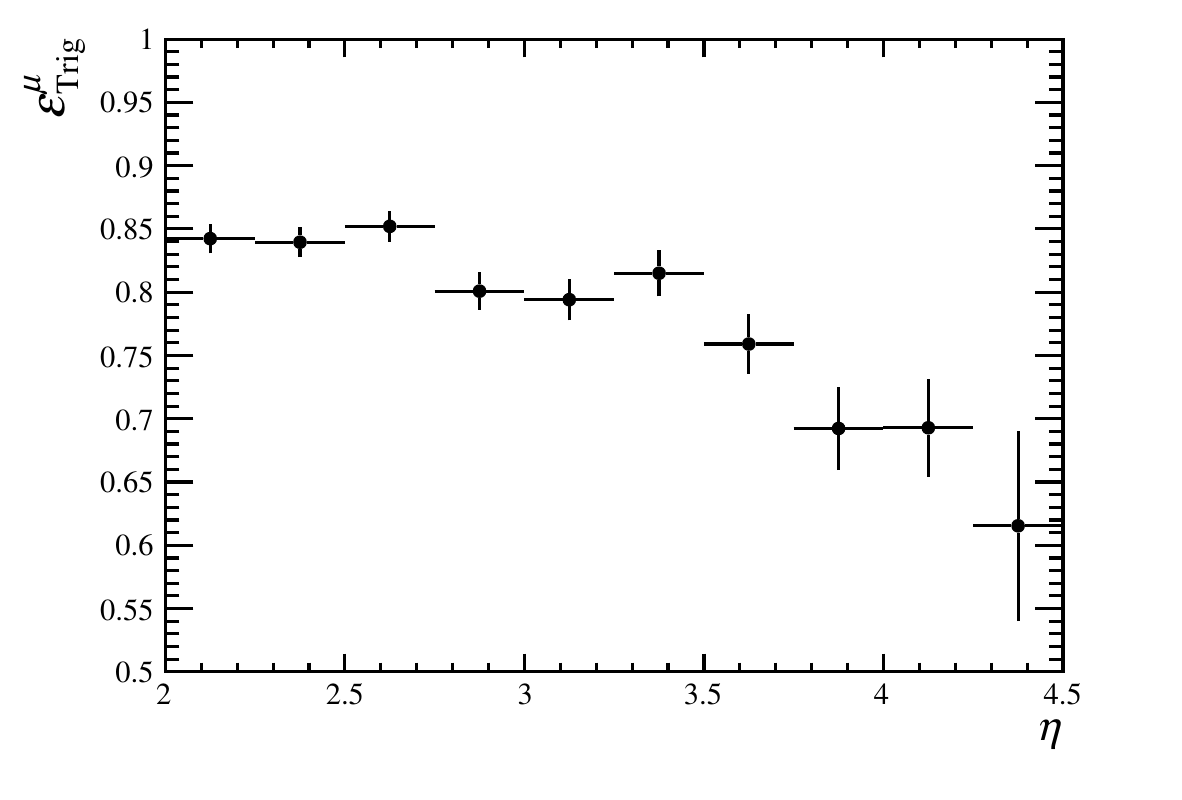}
\caption{Muon tracking, identification and trigger efficiency as a function of pseudorapidity, estimated from the data and simulation at 5.02\tev and 13\tev. }
\label{fig:muon_eff}
\end{center}
\end{figure}

Results of muon efficiencies used in the total \Z boson efficiency calculation are presented in Fig.~\ref{fig:muon_eff} and Table~\ref{tab:muon_eff}.

\begin{table}[tb]
\begin{center}
\caption{The muon tracking, muon identification and trigger efficiency in each pseudorapidity bin.}
\begin{tabular}{cccc}
\hline
$\eta$ & $\varepsilon^{\mu}_{\rm{Track}}\,[\%]$ & $\varepsilon^{\mu}_{\rm{ID}}\,[\%]$ & $\varepsilon^{\mu}_{\rm{Trig}}\,[\%]$  \\ \hline
$[\,$2.00, 2.25$\,]$  &  93.77  &  96.26  &  84.23 \\  
$[\,$2.25, 2.50$\,]$  &  96.82  &  96.30  &  83.95 \\  
$[\,$2.50, 2.75$\,]$  &  97.51  &  95.85  &  85.20 \\  
$[\,$2.75, 3.00$\,]$  &  95.14  &  96.55  &  80.08 \\  
$[\,$3.00, 3.25$\,]$  &  96.26  &  96.73  &  79.41 \\  
$[\,$3.25, 3.50$\,]$  &  98.32  &  96.92  &  81.50 \\  
$[\,$3.50, 3.75$\,]$  &  97.64  &  95.02  &  75.90 \\  
$[\,$3.75, 4.00$\,]$  &  96.00  &  97.08  &  69.23 \\  
$[\,$4.00, 4.25$\,]$  &  96.41  &  97.28  &  69.29 \\  
$[\,$4.25, 4.50$\,]$  &  93.86  &  90.10  &  61.54 \\
\hline
\end{tabular}
\label{tab:muon_eff}
\end{center}
\end{table}

\section{Summary of systematic uncertainties}
\label{app:results_sys}
The summarised systematic uncertainties for single differential cross-sections are shown in 
Tables~\ref{tab:err_y} to~\ref{tab:err_phi}.

\begin{table}[h]
\caption{Systematic uncertainties on the single differential cross-sections in bins of \zy, presented in percentage.
}
\begin{center}
\begin{tabular}{ccccc} \hline
  \multicolumn{1}{c}{\zy}  & Efficiency & Background & FSR & Closure  \\
\hline
%% outputs/Systematic/SystUncer_OneD_Differential_ZRapidity_2017.tex
$[\,$2.000,\, 2.125$\,]$ & 1.42 & 0.04 & 2.10 & 0.45 \\ 
$[\,$2.125,\, 2.250$\,]$ & 1.37 & 0.77 & 0.00 & 0.31 \\ 
$[\,$2.250,\, 2.375$\,]$ & 1.32 & 0.62 & 0.02 & 0.34 \\ 
$[\,$2.375,\, 2.500$\,]$ & 1.31 & 1.47 & 0.21 & 0.67 \\ 
$[\,$2.500,\, 2.625$\,]$ & 1.32 & 1.42 & 0.13 & 1.17 \\ 
$[\,$2.625,\, 2.750$\,]$ & 1.33 & 0.28 & 0.30 & 0.92 \\ 
$[\,$2.750,\, 2.875$\,]$ & 1.36 & 0.52 & 0.02 & 0.89 \\ 
$[\,$2.875,\, 3.000$\,]$ & 1.40 & 0.39 & 0.42 & 1.13 \\ 
$[\,$3.000,\, 3.125$\,]$ & 1.46 & 0.51 & 0.23 & 0.68 \\ 
$[\,$3.125,\, 3.250$\,]$ & 1.56 & 0.00 & 0.50 & 0.58 \\ 
$[\,$3.250,\, 3.375$\,]$ & 1.63 & 0.00 & 0.07 & 1.01 \\ 
$[\,$3.375,\, 3.625$\,]$ & 1.85 & 1.14 & 0.34 & 0.85 \\ 
$[\,$3.625,\, 4.000$\,]$ & 2.23 & 0.00 & 0.21 & 0.80 \\
\hline
\end{tabular}
\label{tab:err_y}
\end{center}
\end{table}

\begin{table}[h]
\caption{Systematic uncertainties on the single differential cross-sections in bins of \zpt, presented in percentage.}
\begin{center}
\begin{tabular}{ccccccc} \hline
 \multicolumn{1}{c}{\zpt~[\gevc]}  & Efficiency & Background & FSR & Closure & Calibration & Migration \\
\hline
%% outputs/Systematic/SystUncer_OneD_Differential_ZPT_2017.tex
$[\,$0.0,\, 2.2$\,]$ & 1.42 & 0.56 & 1.15 & 0.43 & 0.13 & 0.54 \\ 
$[\,$2.2,\, 3.4$\,]$ & 1.41 & 0.50 & 0.64 & 0.37 & 0.40 & 1.73 \\ 
$[\,$3.4,\, 4.6$\,]$ & 1.43 & 0.09 & 0.09 & 0.50 & 0.28 & 1.66 \\ 
$[\,$4.6,\, 5.8$\,]$ & 1.40 & 0.04 & 0.55 & 0.05 & 0.74 & 1.18 \\ 
$[\,$5.8,\, 7.2$\,]$ & 1.43 & 0.25 & 1.27 & 0.12 & 0.91 & 0.36 \\ 
$[\,$7.2,\, 8.7$\,]$ & 1.38 & 0.60 & 1.19 & 2.11 & 0.59 & 0.02 \\ 
$[\,$8.7,\, 10.5$\,]$ & 1.44 & 0.20 & 0.14 & 0.50 & 0.54 & 0.04 \\ 
$[\,$10.5,\, 12.8$\,]$ & 1.39 & 1.34 & 0.48 & 0.97 & 0.31 & 0.03 \\ 
$[\,$12.8,\, 15.4$\,]$ & 1.39 & 2.00 & 1.11 & 0.18 & 0.36 & 0.01 \\ 
$[\,$15.4,\, 19.0$\,]$ & 1.39 & 1.10 & 0.71 & 1.16 & 0.17 & 0.00 \\ 
$[\,$19.0,\, 34.0$\,]$ & 1.37 & 1.09 & 0.09 & 1.97 & 0.10 & 0.00 \\ 
$[\,$34.0,\, 120.0$\,]$ & 1.36 & 1.03 & 0.31 & 2.18 & 0.16 & 0.00 \\ 
\hline
\end{tabular}
\label{tab:err_pt}
\end{center}
\end{table}

\begin{table}[h]
\caption{Systematic uncertainties on the single differential cross-sections in bins of \phistar, presented in percentage.}
\begin{center}
\begin{tabular}{ccccc} \hline
  \multicolumn{1}{c}{\phistar}  & Efficiency & Background & FSR & Closure \\
\hline
%% outputs/Systematic/SystUncer_OneD_Differential_ZPHI_2017.tex
$[\,$0.00,\, 0.01$\,]$ & 1.41 & 1.10 & 0.81 & 0.01 \\ 
$[\,$0.01,\, 0.02$\,]$ & 1.42 & 0.43 & 0.21 & 0.00 \\ 
$[\,$0.02,\, 0.03$\,]$ & 1.41 & 0.14 & 0.36 & 0.12 \\ 
$[\,$0.03,\, 0.05$\,]$ & 1.42 & 0.52 & 0.28 & 0.45 \\ 
$[\,$0.05,\, 0.07$\,]$ & 1.41 & 0.01 & 0.50 & 0.44 \\ 
$[\,$0.07,\, 0.10$\,]$ & 1.39 & 0.29 & 0.24 & 2.22 \\ 
$[\,$0.10,\, 0.15$\,]$ & 1.39 & 1.45 & 0.25 & 0.81 \\ 
$[\,$0.15,\, 0.20$\,]$ & 1.39 & 0.82 & 0.63 & 1.35 \\ 
$[\,$0.20,\, 0.30$\,]$ & 1.38 & 0.77 & 0.97 & 1.00 \\ 
$[\,$0.30,\, 0.60$\,]$ & 1.37 & 1.45 & 0.61 & 2.72 \\ 
$[\,$0.60,\, 1.20$\,]$ & 1.38 & 0.09 & 0.13 & 2.13 \\ 
\hline
\end{tabular}
\label{tab:err_phi}
\end{center}
\end{table}

\newpage

\section{Numerical results of single differential cross-sections}
\label{app:results_cs}
The measured single differential cross-sections in bins of \zy, \zpt and \phistar are presented in Tables~\ref{tab:cen_y} to~\ref{tab:cen_phi}.

%%%%%%%%%%%%%%   uncertainty of  rapidity r%%%%%%%%%%%%%%%%%%%%%%%%%%%%%%%
\begin{table}[h]
\caption{Measured single differential cross-sections in bins of \zy. 
The first uncertainty is statistical, the second systematic, and the third is from the uncertainty on the integrated luminosity.}
\begin{center}
\begin{tabular}{cccccccc}
\hline
  \multicolumn{1}{c}{\zy}  & \multicolumn{7}{c}{$d\sigma(\Zmm)/d{\zy}$ [$\pb$]} \\
\hline
%% outputs/Systematic/RESULTS_OneD_Differential_ZRapidity_2017.tex
$[\,$2.000,\, 2.125$\,]$ & 4.4 &$\pm$& 0.7 &$\pm$& 0.1 &$\pm$& 0.1 \\ 
$[\,$2.125,\, 2.250$\,]$ & 14.6 &$\pm$& 1.2 &$\pm$& 0.2 &$\pm$& 0.3 \\ 
$[\,$2.250,\, 2.375$\,]$ & 24.3 &$\pm$& 1.5 &$\pm$& 0.4 &$\pm$& 0.5 \\ 
$[\,$2.375,\, 2.500$\,]$ & 27.5 &$\pm$& 1.6 &$\pm$& 0.6 &$\pm$& 0.6 \\ 
$[\,$2.500,\, 2.625$\,]$ & 36.1 &$\pm$& 1.8 &$\pm$& 0.8 &$\pm$& 0.7 \\ 
$[\,$2.625,\, 2.750$\,]$ & 39.9 &$\pm$& 2.0 &$\pm$& 0.7 &$\pm$& 0.8 \\ 
$[\,$2.750,\, 2.875$\,]$ & 42.6 &$\pm$& 2.0 &$\pm$& 0.7 &$\pm$& 0.9 \\ 
$[\,$2.875,\, 3.000$\,]$ & 38.8 &$\pm$& 1.9 &$\pm$& 0.7 &$\pm$& 0.8 \\ 
$[\,$3.000,\, 3.125$\,]$ & 30.8 &$\pm$& 1.7 &$\pm$& 0.5 &$\pm$& 0.6 \\ 
$[\,$3.125,\, 3.250$\,]$ & 25.5 &$\pm$& 1.6 &$\pm$& 0.4 &$\pm$& 0.5 \\ 
$[\,$3.250,\, 3.375$\,]$ & 16.0 &$\pm$& 1.3 &$\pm$& 0.3 &$\pm$& 0.3 \\ 
$[\,$3.375,\, 3.625$\,]$ & 6.8 &$\pm$& 0.6 &$\pm$& 0.2 &$\pm$& 0.1 \\ 
$[\,$3.625,\, 4.000$\,]$ & 0.7 &$\pm$& 0.2 &$\pm$& 0.0 &$\pm$& 0.0 \\
\hline
\end{tabular}
\label{tab:cen_y}
\end{center}
\end{table}

%%%%%%%%%%%%%%   uncertainty of  Zpt r%%%%%%%%%%%%%%%%%%%%%%%%%%%%%%%
\begin{table}[h]
\caption{Measured single differential cross-sections in bins of \zpt. 
The first uncertainty is statistical, the second systematic, and the third is due to the luminosity. }
\begin{center}
\begin{tabular}{cccccccc}
\hline
  \multicolumn{1}{c}{\zpt~[\gevc]}  & \multicolumn{7}{c}{$d\sigma(\Zmm)/d{\zpt}$ [$\pb$/(\gevc)]} \\
\hline
%% outputs/Systematic/RESULTS_OneD_Differential_ZPT_2017.tex
$[\,$0.0,\, 2.2$\,]$ & 1.769 &$\pm$& 0.105 &$\pm$& 0.036 &$\pm$& 0.035 \\ 
$[\,$2.2,\, 3.4$\,]$ & 2.77 &$\pm$& 0.17 &$\pm$& 0.07 &$\pm$& 0.06 \\ 
$[\,$3.4,\, 4.6$\,]$ & 3.23 &$\pm$& 0.19 &$\pm$& 0.07 &$\pm$& 0.06 \\ 
$[\,$4.6,\, 5.8$\,]$ & 2.71 &$\pm$& 0.17 &$\pm$& 0.06 &$\pm$& 0.05 \\ 
$[\,$5.8,\, 7.2$\,]$ & 2.27 &$\pm$& 0.14 &$\pm$& 0.05 &$\pm$& 0.05 \\ 
$[\,$7.2,\, 8.7$\,]$ & 2.09 &$\pm$& 0.13 &$\pm$& 0.06 &$\pm$& 0.04 \\ 
$[\,$8.7,\, 10.5$\,]$ & 1.806 &$\pm$& 0.110 &$\pm$& 0.030 &$\pm$& 0.036 \\ 
$[\,$10.5,\, 12.8$\,]$ & 1.234 &$\pm$& 0.076 &$\pm$& 0.028 &$\pm$& 0.025 \\ 
$[\,$12.8,\, 15.4$\,]$ & 0.911 &$\pm$& 0.063 &$\pm$& 0.025 &$\pm$& 0.018 \\ 
$[\,$15.4,\, 19.0$\,]$ & 0.759 &$\pm$& 0.050 &$\pm$& 0.017 &$\pm$& 0.015 \\ 
$[\,$19.0,\, 34.0$\,]$ & 0.328 &$\pm$& 0.016 &$\pm$& 0.009 &$\pm$& 0.007 \\ 
$[\,$34.0,\, 120.0$\,]$ & 0.0325 &$\pm$& 0.0021 &$\pm$& 0.0009 &$\pm$& 0.0007 \\
\hline
\end{tabular}
\label{tab:cen_Zpt}
\end{center}
\end{table}

%%%%%%%%%%%%%%   uncertainty of  phistar%%%%%%%%%%%%%%%%%%%%%%%%%%%%%%%
\begin{table}[h]
\caption{Measured single differential cross-sections in bins of \phistar. 
The first uncertainty is statistical, the second systematic, and the third is due to the luminosity.}
\begin{center}
\begin{tabular}{cccccccc}
\hline
  \multicolumn{1}{c}{\phistar}  & \multicolumn{7}{c}{$d\sigma(\Zmm)/d{\phistar}$ [$\pb$]} \\
\hline
%% outputs/Systematic/RESULTS_OneD_Differential_ZPHI_2017.tex
$[\,$0.00,\, 0.01$\,]$ & 468.5 &$\pm$& 23.8 &$\pm$& 9.2 &$\pm$& 9.4 \\ 
$[\,$0.01,\, 0.02$\,]$ & 442.4 &$\pm$& 23.4 &$\pm$& 6.6 &$\pm$& 8.8 \\ 
$[\,$0.02,\, 0.03$\,]$ & 384.6 &$\pm$& 21.9 &$\pm$& 5.6 &$\pm$& 7.7 \\ 
$[\,$0.03,\, 0.05$\,]$ & 311.6 &$\pm$& 13.8 &$\pm$& 5.0 &$\pm$& 6.2 \\ 
$[\,$0.05,\, 0.07$\,]$ & 212.1 &$\pm$& 11.5 &$\pm$& 3.3 &$\pm$& 4.2 \\ 
$[\,$0.07,\, 0.10$\,]$ & 160.4 &$\pm$& 8.0 &$\pm$& 4.2 &$\pm$& 3.2 \\ 
$[\,$0.10,\, 0.15$\,]$ & 87.63 &$\pm$& 4.51 &$\pm$& 1.91 &$\pm$& 1.75 \\ 
$[\,$0.15,\, 0.20$\,]$ & 45.59 &$\pm$& 3.31 &$\pm$& 1.00 &$\pm$& 0.91 \\ 
$[\,$0.20,\, 0.30$\,]$ & 24.21 &$\pm$& 1.70 &$\pm$& 0.51 &$\pm$& 0.48 \\ 
$[\,$0.30,\, 0.60$\,]$ & 6.30 &$\pm$& 0.49 &$\pm$& 0.22 &$\pm$& 0.13 \\ 
$[\,$0.60,\, 1.20$\,]$ & 0.571 &$\pm$& 0.108 &$\pm$& 0.015 &$\pm$& 0.011 \\ 
\hline
\end{tabular}
\label{tab:cen_phi}
\end{center}
\end{table}

\section{Correlation matrices}
\label{app:correlation}

\begin{figure}[h]
\begin{center}
\includegraphics[width=0.45\textwidth]{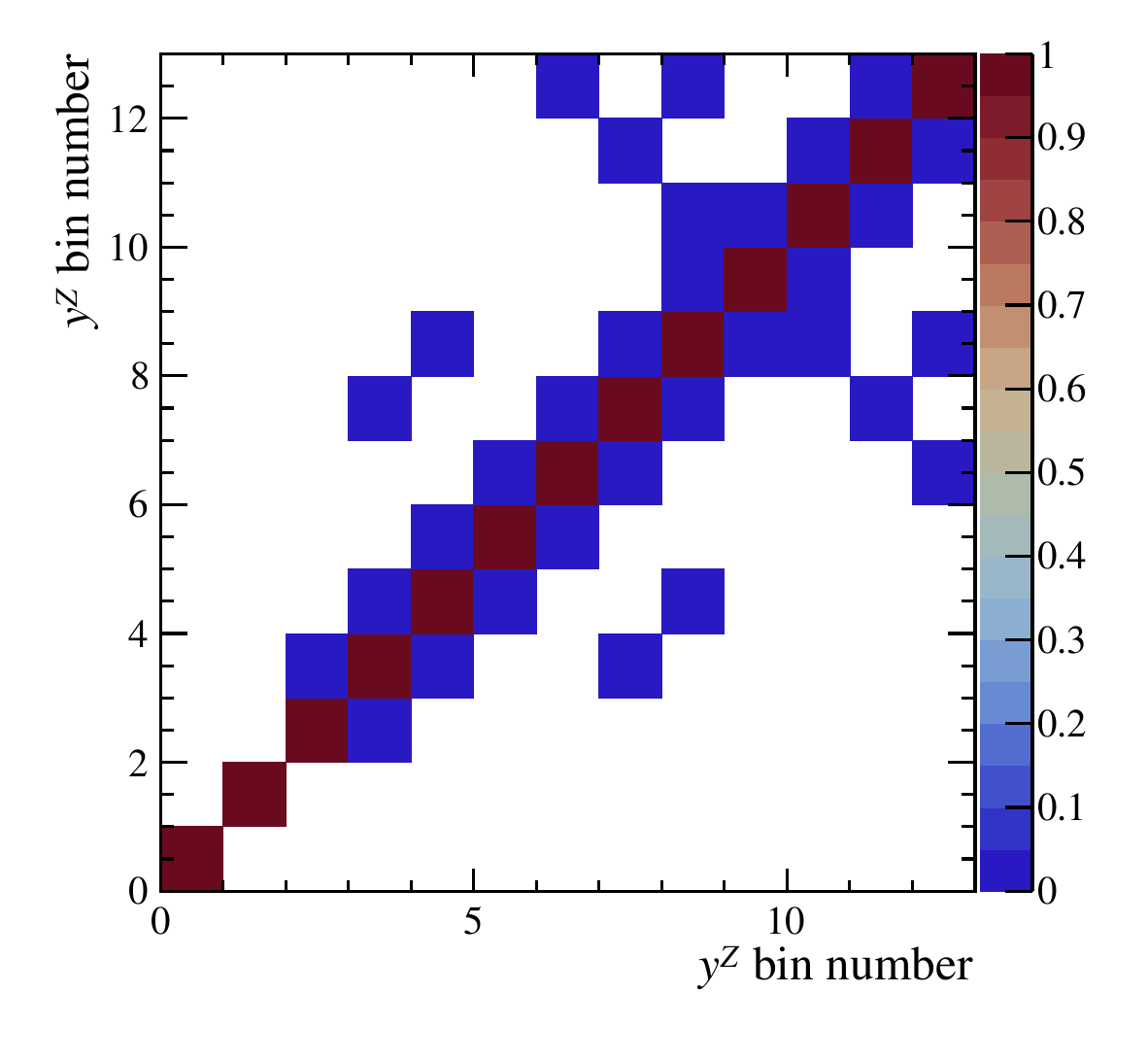}
\includegraphics[width=0.45\textwidth]{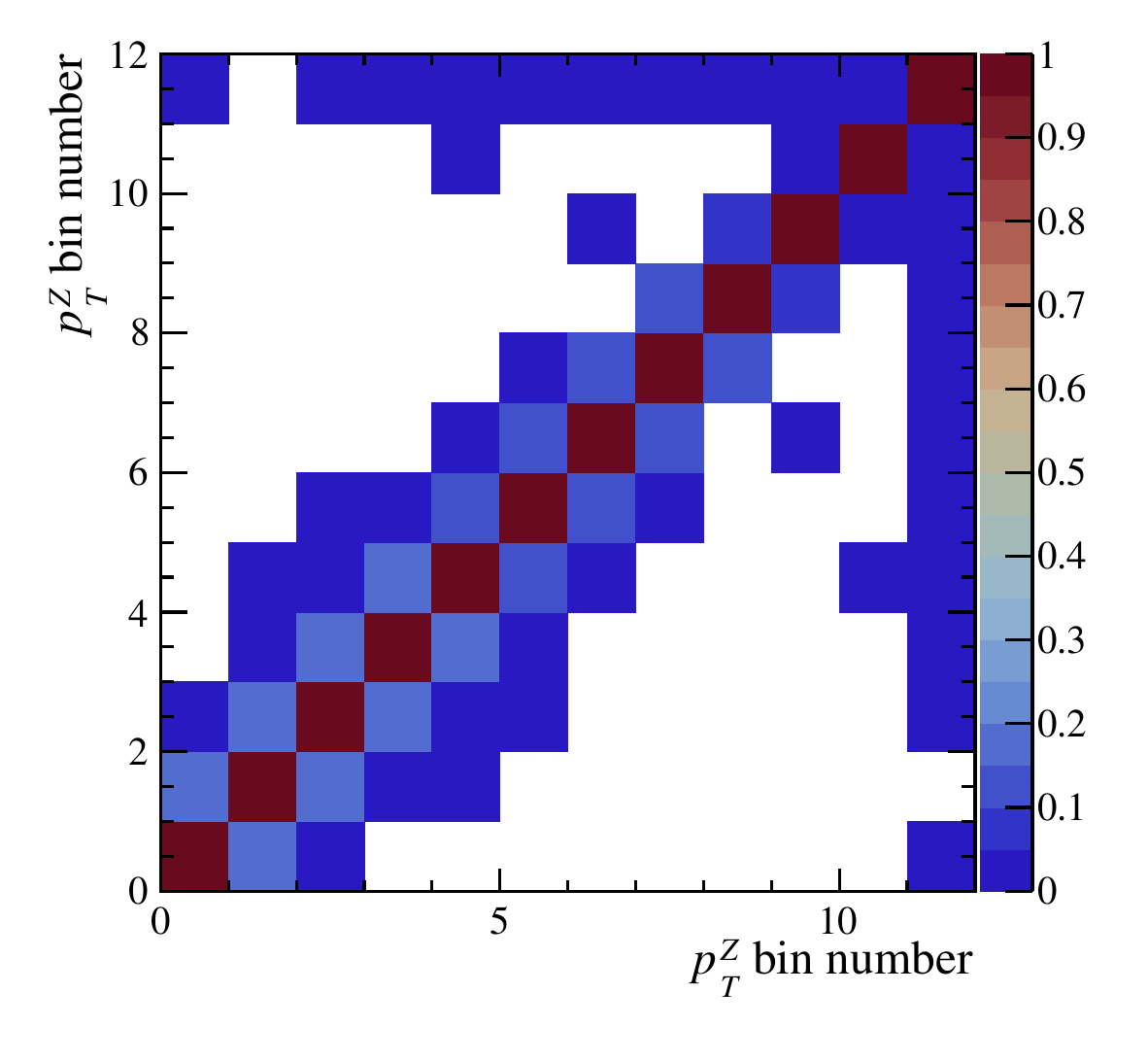}
\includegraphics[width=0.45\textwidth]{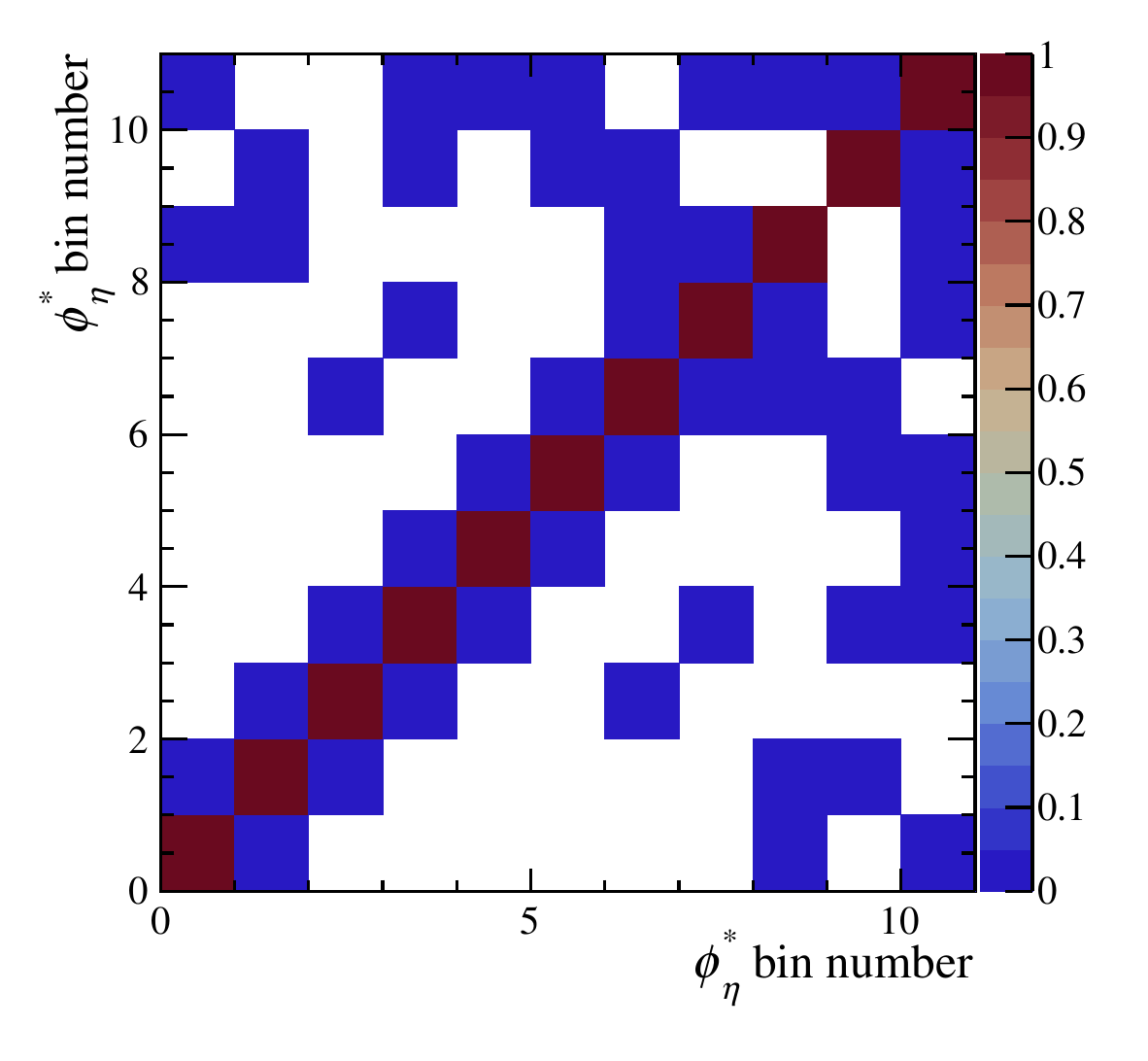}
\caption{Statistical correlation matrices for the differential
cross-section measurements as functions of (top-left) \zy, (top-right) \zpt and (bottom) \phistar.}

\label{fig:corr_stat_1D}
\end{center}
\end{figure}

\begin{figure}[h]
\begin{center}
\includegraphics[width=0.45\textwidth]{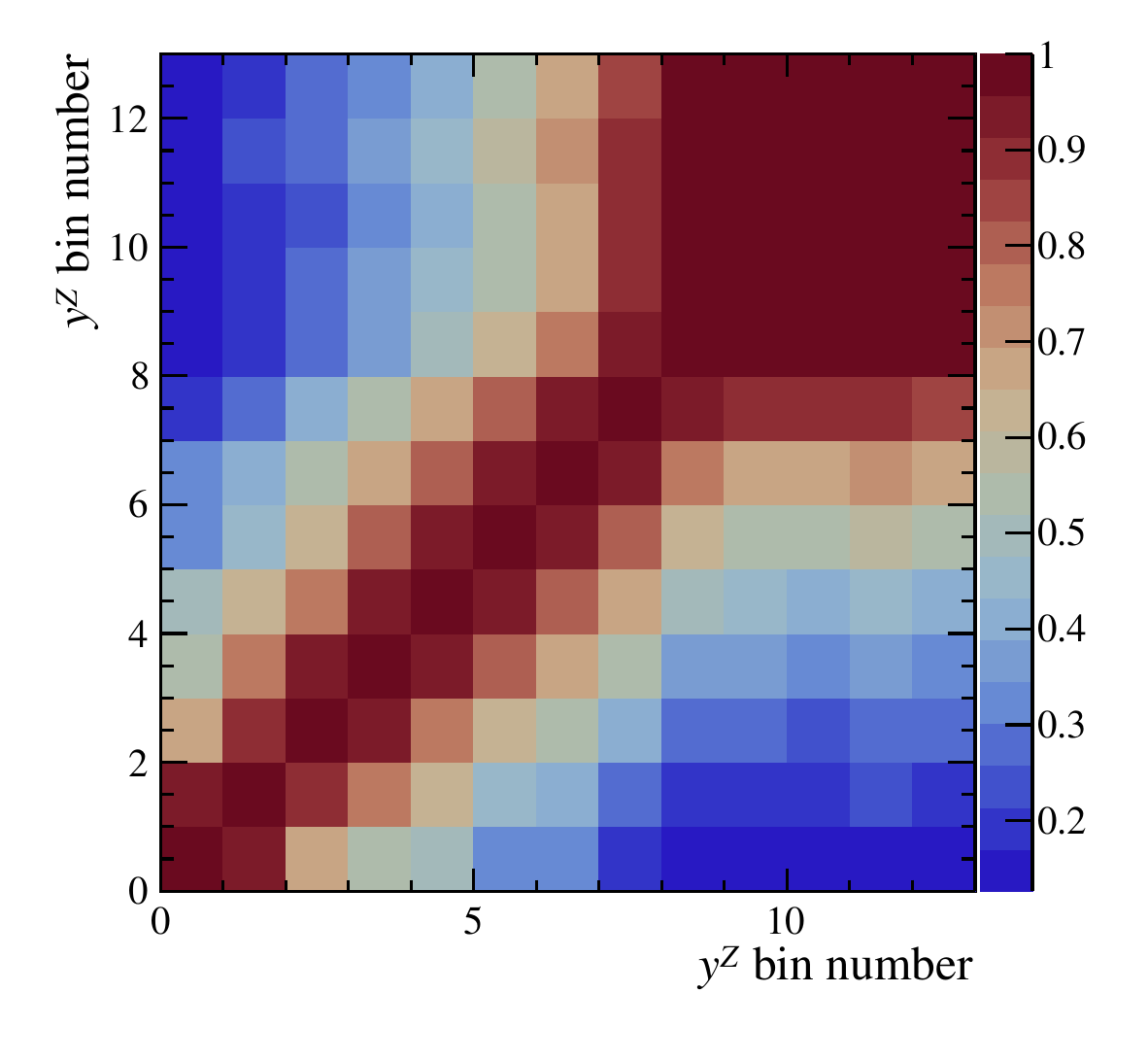}
\includegraphics[width=0.45\textwidth]{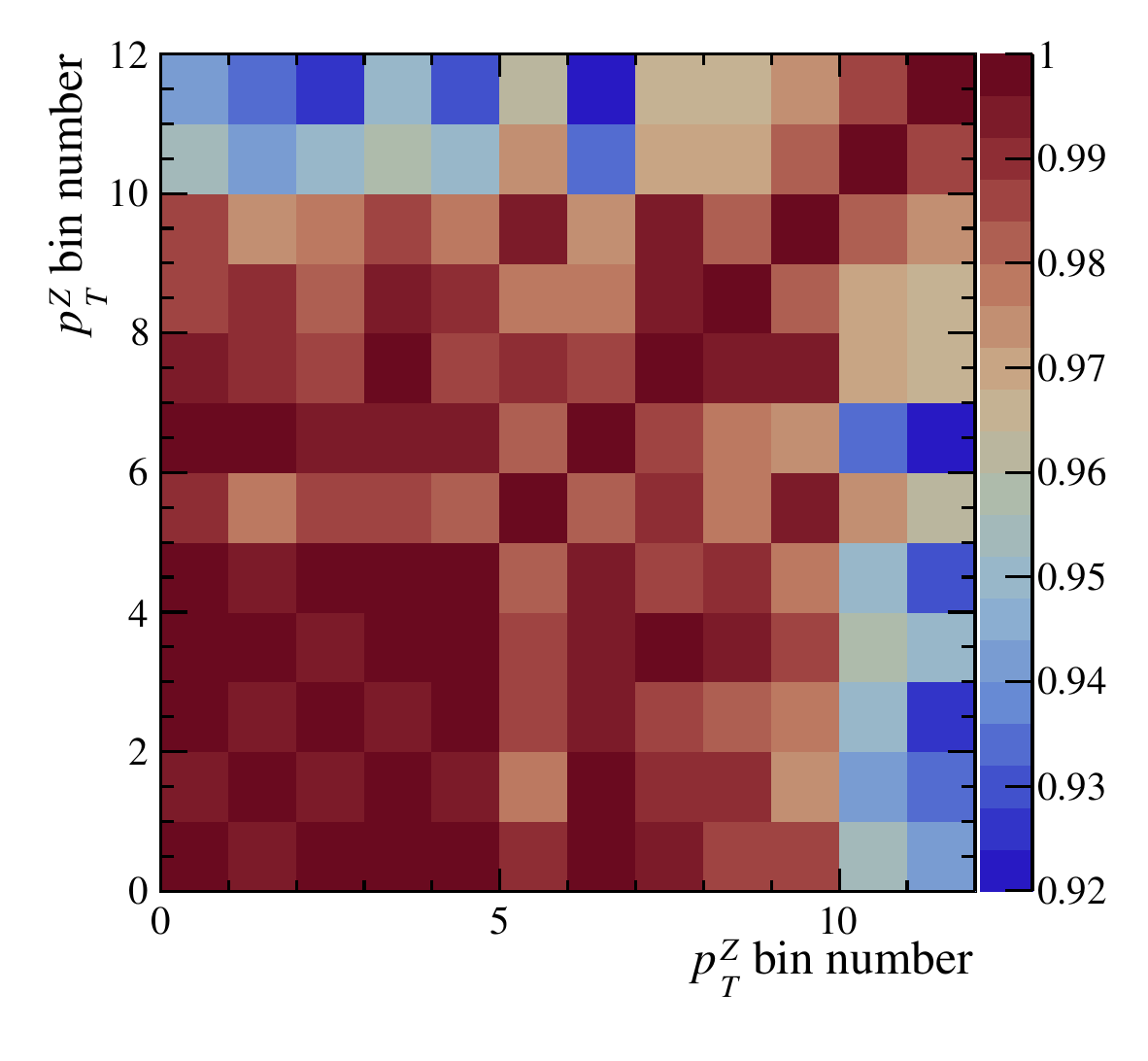}
\includegraphics[width=0.45\textwidth]{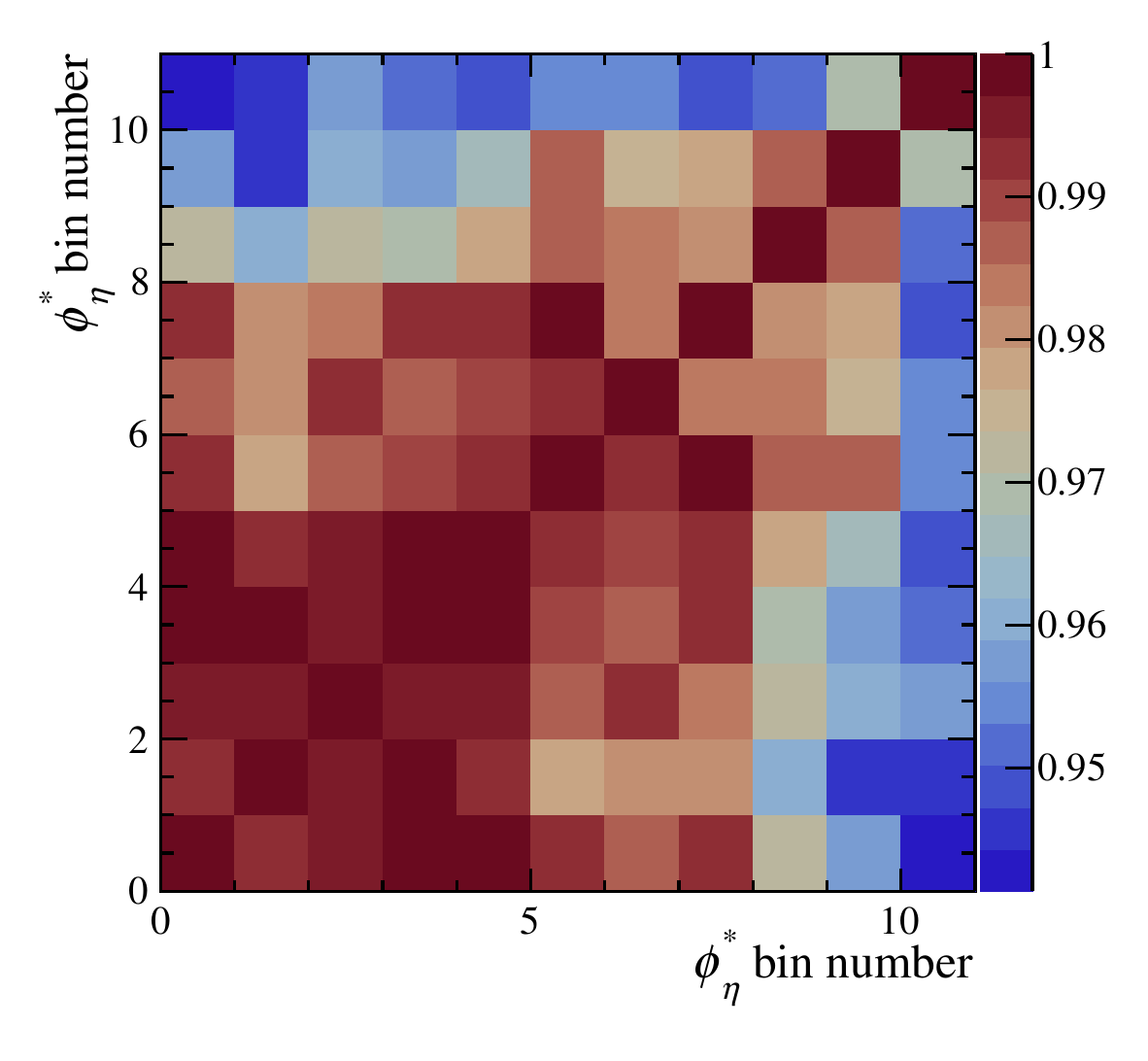}
\caption{Correlation matrices for the efficiency uncertainties as functions of (top-left) \zy, (top-right) \zpt and (bottom) \phistar.}

\label{fig:corr_syst_1D}
\end{center}
\end{figure}

The calculated statistical correlation matrices are shown in Fig.~\ref{fig:corr_stat_1D} and presented in Tables~\ref{tab:LPer_ystat} to~\ref{tab:LPer_phistat}.
The correlation matrices for the efficiency uncertainty
are shown in Fig.~\ref{fig:corr_syst_1D} for single differential cross-section measurements, and presented in Tables~\ref{tab:LPer_ysyst} to~\ref{tab:LPer_phisyst}. 

%%%%%%%%%%%%%%%%%%%%%%%%%%%%%%%%%%%%%%%%%%%%%%%%%%%%%%%%%%%%%%%%%%%%%%%%%%%%%
\begin{sidewaystable}[h]
\caption{Statistical correlation matrix for the one-dimensional \zy measurement.} 
\centering 
\begin{tabular}{c|cccccccccccccccc} 
%% outputs/Systematic/RESULTS_Stat_1D_Correlation_ZRapidity.tex
Bin & 1 & 2 & 3 & 4 & 5 & 6 & 7 & 8 & 9 & 10 & 11 & 12 & 13   \\ \hline
1  & 1.00 &      &      &      &      &      &      &      &      &      &      &      &      \\ 
2  & 0.00 & 1.00 &      &      &      &      &      &      &      &      &      &      &      \\ 
3  & 0.00 & 0.00 & 1.00 &      &      &      &      &      &      &      &      &      &      \\ 
4  & 0.00 & 0.00 & 0.00 & 1.00 &      &      &      &      &      &      &      &      &      \\ 
5  & 0.00 & 0.00 & 0.00 & 0.01 & 1.00 &      &      &      &      &      &      &      &      \\ 
6  & 0.00 & 0.00 & 0.00 & 0.00 & 0.01 & 1.00 &      &      &      &      &      &      &      \\ 
7  & 0.00 & 0.00 & 0.00 & 0.00 & 0.00 & 0.01 & 1.00 &      &      &      &      &      &      \\ 
8  & 0.00 & 0.00 & 0.00 & 0.00 & 0.00 & 0.00 & 0.01 & 1.00 &      &      &      &      &      \\ 
9  & 0.00 & 0.00 & 0.00 & 0.00 & 0.00 & 0.00 & 0.00 & 0.01 & 1.00 &      &      &      &      \\ 
10  & 0.00 & 0.00 & 0.00 & 0.00 & 0.00 & 0.00 & 0.00 & 0.00 & 0.01 & 1.00 &      &      &      \\ 
11  & 0.00 & 0.00 & 0.00 & 0.00 & 0.00 & 0.00 & 0.00 & 0.00 & 0.00 & 0.01 & 1.00 &      &      \\ 
12  & 0.00 & 0.00 & 0.00 & 0.00 & 0.00 & 0.00 & 0.00 & 0.00 & 0.00 & 0.00 & 0.02 & 1.00 &      \\ 
13  & 0.00 & 0.00 & 0.00 & 0.00 & 0.00 & 0.00 & 0.00 & 0.00 & 0.00 & 0.00 & 0.00 & 0.00 & 1.00   \\
\hline 
\end{tabular}
\label{tab:LPer_ystat}
\end{sidewaystable}
%%%%%%%%%%%%%%%%%%%%%%%%%%%%%%%%%%%%%%%%%%%%%%%%%%%%%%%%%%%%%%%%%%%%%%%%%%%%%

\begin{sidewaystable}[h]
\caption{Statistical correlation matrix for the one-dimensional \zpt measurement.} 
\centering 
\begin{tabular}{c|cccccccccccccccc} 
%% outputs/Systematic/RESULTS_Stat_1D_Correlation_ZPT.tex
Bin & 1 & 2 & 3 & 4 & 5 & 6 & 7 & 8 & 9 & 10 & 11 & 12     \\ \hline
1  & 1.00 &      &      &      &      &      &      &      &      &      &      &      \\ 
2  & 0.16 & 1.00 &      &      &      &      &      &      &      &      &      &      \\ 
3  & 0.00 & 0.18 & 1.00 &      &      &      &      &      &      &      &      &      \\ 
4  & 0.00 & 0.01 & 0.18 & 1.00 &      &      &      &      &      &      &      &      \\ 
5  & 0.00 & 0.00 & 0.01 & 0.18 & 1.00 &      &      &      &      &      &      &      \\ 
6  & 0.00 & 0.00 & 0.00 & 0.01 & 0.14 & 1.00 &      &      &      &      &      &      \\ 
7  & 0.00 & 0.00 & 0.00 & 0.00 & 0.01 & 0.12 & 1.00 &      &      &      &      &      \\ 
8  & 0.00 & 0.00 & 0.00 & 0.00 & 0.00 & 0.00 & 0.10 & 1.00 &      &      &      &      \\ 
9  & 0.00 & 0.00 & 0.00 & 0.00 & 0.00 & 0.00 & 0.00 & 0.10 & 1.00 &      &      &      \\ 
10  & 0.00 & 0.00 & 0.00 & 0.00 & 0.00 & 0.00 & 0.00 & 0.00 & 0.05 & 1.00 &      &      \\ 
11  & 0.00 & 0.00 & 0.00 & 0.00 & 0.00 & 0.00 & 0.00 & 0.00 & 0.00 & 0.04 & 1.00 &      \\ 
12  & 0.00 & 0.00 & 0.00 & 0.00 & 0.00 & 0.01 & 0.00 & 0.00 & 0.00 & 0.01 & 0.01 & 1.00   \\ 
\hline 
\end{tabular}
\label{tab:LPer_ptstat}
\end{sidewaystable}

%%%%%%%%%%%%%%%%%%%%%%%%%%%%%%%%%%%%%%%%%%%%%%%%%%%%%%%%%%%%%%%%%%%%%%%%%%%%%
\begin{sidewaystable}[h]
\caption{Statistical correlation matrix for the one-dimensional \phistar measurement.} 
\centering 
\begin{tabular}{c|cccccccccccccccc} 
%% outputs/Systematic/RESULTS_Stat_1D_Correlation_ZPHI.tex
Bin & 1 & 2 & 3 & 4 & 5 & 6 & 7 & 8 & 9 & 10 & 11    \\ \hline
1  & 1.00 &      &      &      &      &      &      &      &      &      &      \\ 
2  & 0.01 & 1.00 &      &      &      &      &      &      &      &      &      \\ 
3  & 0.00 & 0.01 & 1.00 &      &      &      &      &      &      &      &      \\ 
4  & 0.00 & 0.00 & 0.01 & 1.00 &      &      &      &      &      &      &      \\ 
5  & 0.00 & 0.00 & 0.00 & 0.01 & 1.00 &      &      &      &      &      &      \\ 
6  & 0.00 & 0.00 & 0.00 & 0.00 & 0.00 & 1.00 &      &      &      &      &      \\ 
7  & 0.00 & 0.00 & 0.00 & 0.00 & 0.00 & 0.00 & 1.00 &      &      &      &      \\ 
8  & 0.00 & 0.00 & 0.00 & 0.00 & 0.00 & 0.00 & 0.00 & 1.00 &      &      &      \\ 
9  & 0.00 & 0.00 & 0.00 & 0.00 & 0.00 & 0.00 & 0.00 & 0.00 & 1.00 &      &      \\ 
10  & 0.00 & 0.00 & 0.00 & 0.00 & 0.00 & 0.00 & 0.00 & 0.00 & 0.00 & 1.00 &      \\ 
11  & 0.00 & 0.00 & 0.00 & 0.00 & 0.00 & 0.00 & 0.00 & 0.00 & 0.00 & 0.00 & 1.00   \\
\hline
\end{tabular}
\label{tab:LPer_phistat}
\end{sidewaystable}

%%%%%%%%%%%%%%%%%%%%%%%%%%%%%%%%%%%%%%%%%%%%%%%%%%%%%%%%%%%%%%%%%%%%%%%%%%%%%
%%%%%%%%%%%%%%%%%%%%%%%%%%%%%%%%%%%%%%%%%%%%%%%%%%%%%%%%%%%%%%%%%%%%%%%%%%%%%

\begin{sidewaystable}[h]
\caption{Correlation matrix for the efficiency uncertainty of the one-dimensional \zy measurement.} 
\centering 
\begin{tabular}{c|cccccccccccccccc} 
%% outputs/Systematic/RESULTS_Syst_1D_Correlation_ZRapidity.tex
Bin & 1 & 2 & 3 & 4 & 5 & 6 & 7 & 8 & 9 & 10 & 11 & 12 & 13   \\ \hline
1  & 1.00 &      &      &      &      &      &      &      &      &      &      &      &      \\ 
2  & 0.92 & 1.00 &      &      &      &      &      &      &      &      &      &      &      \\ 
3  & 0.68 & 0.89 & 1.00 &      &      &      &      &      &      &      &      &      &      \\ 
4  & 0.56 & 0.74 & 0.93 & 1.00 &      &      &      &      &      &      &      &      &      \\ 
5  & 0.49 & 0.62 & 0.77 & 0.92 & 1.00 &      &      &      &      &      &      &      &      \\ 
6  & 0.34 & 0.47 & 0.65 & 0.81 & 0.95 & 1.00 &      &      &      &      &      &      &      \\ 
7  & 0.31 & 0.41 & 0.54 & 0.67 & 0.82 & 0.94 & 1.00 &      &      &      &      &      &      \\ 
8  & 0.19 & 0.28 & 0.40 & 0.53 & 0.69 & 0.81 & 0.93 & 1.00 &      &      &      &      &      \\ 
9  & 0.14 & 0.21 & 0.29 & 0.38 & 0.51 & 0.63 & 0.78 & 0.94 & 1.00 &      &      &      &      \\ 
10  & 0.13 & 0.19 & 0.27 & 0.35 & 0.44 & 0.54 & 0.69 & 0.87 & 0.99 & 1.00 &      &      &      \\ 
11  & 0.13 & 0.18 & 0.25 & 0.33 & 0.42 & 0.52 & 0.68 & 0.87 & 0.98 & 1.00 & 1.00 &      &      \\ 
12  & 0.16 & 0.22 & 0.29 & 0.37 & 0.46 & 0.57 & 0.72 & 0.89 & 0.99 & 1.00 & 1.00 & 1.00 &      \\ 
13  & 0.16 & 0.21 & 0.27 & 0.34 & 0.42 & 0.52 & 0.68 & 0.87 & 0.98 & 1.00 & 1.00 & 1.00 & 1.00   \\ 
\hline
\end{tabular}
\label{tab:LPer_ysyst}
\end{sidewaystable}

%%%%%%%%%%%%%%%%%%%%%%%%%%%%%%%%%%%%%%%%%%%%%%%%%%%%%%%%%%%%%%%%%%%%%%%%%%%%%
\begin{sidewaystable}[h]
\caption{Correlation matrix for the efficiency uncertainty of the one-dimensional \zpt measurement.} 
\centering 
\begin{tabular}{c|cccccccccccccccc}
%%  outputs/Systematic/RESULTS_Syst_1D_Correlation_ZPT.tex
Bin & 1 & 2 & 3 & 4 & 5 & 6 & 7 & 8 & 9 & 10 & 11 & 12    \\ \hline
1  & 1.00 &      &      &      &      &      &      &      &      &      &      &      \\ 
2  & 1.00 & 1.00 &      &      &      &      &      &      &      &      &      &      \\ 
3  & 1.00 & 0.99 & 1.00 &      &      &      &      &      &      &      &      &      \\ 
4  & 1.00 & 1.00 & 1.00 & 1.00 &      &      &      &      &      &      &      &      \\ 
5  & 1.00 & 1.00 & 1.00 & 1.00 & 1.00 &      &      &      &      &      &      &      \\ 
6  & 0.99 & 0.98 & 0.99 & 0.99 & 0.98 & 1.00 &      &      &      &      &      &      \\ 
7  & 1.00 & 1.00 & 1.00 & 1.00 & 1.00 & 0.98 & 1.00 &      &      &      &      &      \\ 
8  & 0.99 & 0.99 & 0.99 & 1.00 & 0.99 & 0.99 & 0.99 & 1.00 &      &      &      &      \\ 
9  & 0.99 & 0.99 & 0.98 & 0.99 & 0.99 & 0.98 & 0.98 & 0.99 & 1.00 &      &      &      \\ 
10  & 0.98 & 0.98 & 0.98 & 0.99 & 0.98 & 0.99 & 0.97 & 0.99 & 0.98 & 1.00 &      &      \\ 
11  & 0.96 & 0.94 & 0.95 & 0.96 & 0.95 & 0.97 & 0.93 & 0.97 & 0.97 & 0.98 & 1.00 &      \\ 
12  & 0.94 & 0.93 & 0.93 & 0.95 & 0.93 & 0.96 & 0.92 & 0.97 & 0.97 & 0.98 & 0.99 & 1.00   \\
\hline 
\end{tabular}
\label{tab:LPer_ptsyst}
\end{sidewaystable}

\begin{sidewaystable}[h]
\caption{Correlation matrix for the efficiency uncertainty of the one-dimensional \phistar measurement.} 
\centering 
\begin{tabular}{c|cccccccccccccccccc}
%%  outputs/Systematic/RESULTS_Syst_1D_Correlation_ZPHI.tex
Bin & 1 & 2 & 3 & 4 & 5 & 6 & 7 & 8 & 9 & 10 & 11   \\ \hline
1  & 1.00 &      &      &      &      &      &      &      &      &      &      \\ 
2  & 0.99 & 1.00 &      &      &      &      &      &      &      &      &      \\ 
3  & 0.99 & 1.00 & 1.00 &      &      &      &      &      &      &      &      \\ 
4  & 1.00 & 1.00 & 1.00 & 1.00 &      &      &      &      &      &      &      \\ 
5  & 1.00 & 0.99 & 0.99 & 1.00 & 1.00 &      &      &      &      &      &      \\ 
6  & 0.99 & 0.98 & 0.99 & 0.99 & 0.99 & 1.00 &      &      &      &      &      \\ 
7  & 0.99 & 0.98 & 0.99 & 0.99 & 0.99 & 0.99 & 1.00 &      &      &      &      \\ 
8  & 0.99 & 0.98 & 0.98 & 0.99 & 0.99 & 1.00 & 0.98 & 1.00 &      &      &      \\ 
9  & 0.97 & 0.96 & 0.97 & 0.97 & 0.98 & 0.99 & 0.98 & 0.98 & 1.00 &      &      \\ 
10  & 0.96 & 0.95 & 0.96 & 0.96 & 0.97 & 0.99 & 0.98 & 0.98 & 0.99 & 1.00 &      \\ 
11  & 0.94 & 0.95 & 0.96 & 0.95 & 0.95 & 0.96 & 0.95 & 0.95 & 0.95 & 0.97 & 1.00   \\
\hline 
\end{tabular}
\label{tab:LPer_phisyst}
\end{sidewaystable}
%%%%%%%%%%%%%%%%%%%%%%%%%%%%%%%%%%%%%%%%%%%%%%%%%%%%%%%%%%%%%%%%%%%%%%%%%%%%%

\clearpage

\clearpage

\addcontentsline{toc}{section}{References}
\bibliographystyle{LHCb}
\bibliography{main,standard,LHCb-PAPER,LHCb-CONF,LHCb-DP,LHCb-TDR}

%%%%%%% Authorship
\newpage
% LHCb collaboration author list
% Data extracted on August 9th, 2023 at 9:47am for paper reference LHCb-PAPER-2023-010
\centerline
{\large\bf LHCb collaboration}
\begin
{flushleft}
\small
R.~Aaij$^{32}$\lhcborcid{0000-0003-0533-1952},
A.S.W.~Abdelmotteleb$^{51}$\lhcborcid{0000-0001-7905-0542},
C.~Abellan~Beteta$^{45}$,
F.~Abudin{\'e}n$^{51}$\lhcborcid{0000-0002-6737-3528},
T.~Ackernley$^{55}$\lhcborcid{0000-0002-5951-3498},
B.~Adeva$^{41}$\lhcborcid{0000-0001-9756-3712},
M.~Adinolfi$^{49}$\lhcborcid{0000-0002-1326-1264},
P.~Adlarson$^{77}$\lhcborcid{0000-0001-6280-3851},
H.~Afsharnia$^{9}$,
C.~Agapopoulou$^{43}$\lhcborcid{0000-0002-2368-0147}
% ...

{\footnotesize \it

% ...
$^{9}$Universit{\'e} Clermont Auvergne, CNRS/IN2P3, LPC, Clermont-Ferrand, France\\
% ...
$^{32}$Nikhef National Institute for Subatomic Physics, Amsterdam, Netherlands\\
% ...
$^{41}$Instituto Galego de F{\'\i}sica de Altas Enerx{\'\i}as (IGFAE), Universidade de Santiago de Compostela, Santiago de Compostela, Spain\\
% ...
$^{43}$European Organization for Nuclear Research (CERN), Geneva, Switzerland\\
% ...
$^{45}$Physik-Institut, Universit{\"a}t Z{\"u}rich, Z{\"u}rich, Switzerland\\
% ...
$^{49}$H.H. Wills Physics Laboratory, University of Bristol, Bristol, United Kingdom\\
% ...
$^{51}$Department of Physics, University of Warwick, Coventry, United Kingdom\\
% ...
$^{54}$School of Physics and Astronomy, University of Glasgow, Glasgow, United Kingdom\\
$^{55}$Oliver Lodge Laboratory, University of Liverpool, Liverpool, United Kingdom\\
% ...
$^{77}$Department of Physics and Astronomy, Uppsala University, Uppsala, Sweden, associated to $^{54}$\\
% ...
}
\end{flushleft}

\end{document}